\documentclass[longauth]{aa}	     

\usepackage{graphicx}
\usepackage{txfonts}
\usepackage{natbib}
\usepackage{color}
\usepackage{textcomp}
\usepackage[ ]{hyperref}

\begin{document} 

   \title{J-PLUS: measuring ${\rm H}\alpha$ emission line fluxes in the nearby universe}  

   \author{R. Logroño-García\inst{1*} \and G. Vilella-Rojo\inst{1} \and C. López-Sanjuan\inst{2} \and J. Varela\inst{2} \and K. Viironen\inst{2} \and D.~J. Muniesa\inst{1} \and A.~J. Cenarro\inst{2} \and D. Cristóbal-Hornillos\inst{2} \and A. Ederoclite\inst{2} \and A. Marín-Franch\inst{2} \and M. Moles\inst{2} \and H. Vázquez Ramió\inst{1} \and S. Bonoli\inst{2} \and L.~A. Díaz-García\inst{1} \and A. Orsi\inst{1} \and I. San Roman\inst{1} \and S. Akras\inst{3} \and A.~L. Chies-Santos\inst{4} \and P.~R.~T. Coelho\inst{5} \and S. Daflon\inst{3} \and M.~V. Costa-Duarte\inst{5} \and R. Dupke\inst{3,6} \and L. Galbany\inst{7} \and R.~M. González Delgado\inst{8} \and J.~A. Hernandez-Jimenez\inst{5} \and R. Lopes de Oliveira\inst{9,3,10,11} \and C. Mendes de Oliveira\inst{5} \and I. Oteo\inst{11,12} \and D.~R. Gon\c{c}alves\inst{13} \and M. Sánchez-Portal\inst{14} \and  L. Schmidtobreick\inst{15} \and  L. Sodré Jr.\inst{5}}
   
   \institute{
   Centro de Estudios de Física del Cosmos de Aragón (CEFCA), Plaza San Juan 1, 44001 Teruel, Spain 
   \and
   Centro de Estudios de Física del Cosmos de Aragón (CEFCA), Unidad Asociada al CSIC, Plaza San Juan 1, 44001 Teruel, Spain 
   \and 
   Observatório Nacional/MCTIC, Rua Gen. José Cristino, 77, 20921-400, Rio de Janeiro, Brazil 
   \and 
   Departamento de Astronomia, Instituto de Física, Universidade Federal do Rio Grande do Sul, Porto Alegre, R.S, Brazil 
   \and 
   Universidade de Sao Paulo, Instituto de Astronomia, Geofísica e Ciências Atmosféricas, Rua do Matao, 1226 Sao Paulo, SP 05508–900, Brazil
   \and
   Dept. of Astronomy, University of Michigan, Ann Arbor, MI 48109-1107, USA
   \and
   PITT PACC, Department of Physics and Astronomy, University of Pittsburgh, Pittsburgh, PA 15260, USA
   \and
   IAA (CSIC), Glorieta de la Astronomía s/n, 18008 Granada, Spain
   \and
   Universidade Federal de Sergipe, Departamento de Física, Av. Marechal Rondon, S/N, 49000-000 Sao Cristóvao, SE, Brazil
   \and
   X-ray Astrophysics Laboratory, NASA Goddard Space Flight Center, Greenbelt, MD 20771, USA
   \and
   Department of Physics, University of Maryland, Baltimore County, 1000 Hilltop Circle, Baltimore, MD 21250, USA
   \and
   Institute for Astronomy, University of Edinburgh, Royal Observatory, Blackford Hill, Edinburgh EH9 3HJ, U.K.
   \and 
   European Southern Observatory, Karl-Schwarzschild-Str.  2, 85748 Garching, Germany
   \and
   Observatório do Valongo, Universidade Federal do Rio de Janeiro, Ladeira Pedro Antonio 43, 20080-090 Rio de Janeiro, Brazil 
   \and
   European Space Astronomy Centre (ESAC)/ESA, P.O. Box 78, 28690 Villanueva de la Canada, Madrid, Spain
   \and
   European Southern Observatory, Casilla 19001, Santiago 19, Chile
   \\
   \email{*rlgarcia@cefca.es}}
   \date{Received: December 18, 2017}

\abstract  
   {In the present paper we aim to validate a methodology designed to extract the ${\rm H}\alpha$ emission line flux from J-PLUS photometric data. J-PLUS is a multi narrow-band filter survey carried out with the 2 deg$^{2}$ field of view T80Cam camera, mounted on the JAST/T80 telescope in the OAJ, Teruel, Spain. The information of the twelve J-PLUS bands, including the $J0660$ narrow-band filter located at rest-frame ${\rm H}\alpha$, is used over 42 deg$^2$ to extract de-reddened and [NII] decontaminated ${\rm H}\alpha$ emission line fluxes of 46 star-forming regions with previous SDSS and/or CALIFA spectroscopic information. The agreement of the inferred J-PLUS photometric ${\rm H}\alpha$ fluxes and those obtained with spectroscopic data is remarkable, with a median comparison ratio $\mathcal{R}=F^{\rm J-PLUS}_{\rm H\alpha}/F^{\rm spec}_{\rm H\alpha} = 1.05 \pm 0.25$. This demonstrates that it is possible to retrieve reliable ${\rm H}\alpha$ emission line fluxes from J-PLUS photometric data. With an expected area of thousands of square degrees upon completion, the J-PLUS dataset will allow the study of several star formation science cases in the nearby universe, as the spatially resolved star formation rate of nearby galaxies at $\rm z \le 0.015$, and how it is influenced by the environment, morphology or nuclear activity. As an illustrative example, the close pair of interacting galaxies NGC3994 and NGC3995 is analyzed, finding an enhancement of the star formation rate not only in the center, but also in outer parts of the disk of NGC3994.}

   \keywords{methods: data analysis -- techniques: photometric -- galaxies: star formation -- galaxies: low z}

\maketitle

\section{Introduction}

The formation and evolution of galaxies is greatly influenced by the rate at which the available gas is converted into stars, namely, the star formation rate (SFR). Characterizing the SFR of galaxies at different epochs, and its relation with environment, morphology, or nuclear activity among others, gives us a general picture of how galaxies form and evolve across cosmic time.

The physical processes occuring at the formation phase of stars, yield observational hints in different wavelength ranges that can be used as tracers of star formation \citep{kenSFR,calzetti2}. Stars are born in molecular gas clouds. In these clouds, the newborn stars ionize the surrounding hydrogen with their ultraviolet emission. The recombination of hydrogen after this, causes the emission of photons with wavelengths in the hydrogen spectral series, as the ones with ${\rm H}\alpha$ wavelength (\textit{n=3} to \textit{n=2} transition). The fact that ${\rm H}\alpha$ rest-frame wavelength lies in the optical range of the spectrum and is less affected by both dust and atmospheric extinction than other tracers, make it an easy detectable emission line, hence a suitable tracer to study the SFR in the nearby universe \citep[e.g.][]{kenicut,catalan}.

Large spectroscopic and photometric surveys, such as the Sloan Digital Sky Survey (SDSS, \citealt{sloan}), have revolutionized astrophysics in many fields, but can only deliver a limited view of the star formation activity in the local universe, given the small field of view (FoV) in the case of spectroscopic surveys, and the low spectral resolution in the case of photometric surveys. In the last years, integral field spectroscopic (IFS) surveys have overcome these problems with the use of larger FoVs with a fully spectral coverage (e.g SAURON, \citealt{SAURON}; ATLAS$^{\rm 3D}$, \citealt{Cappellari}; CALIFA, \citealt{CALIFA}; SAMI, \citealt{Croom}; VENGA, \citealt{VIRUS}; and MaNGA, \citealt{Bundy}). However, they still have limitations, such as the lack of a large contiguous observed area to trace the environment of nearby galaxies. They also suffer from aperture selection effects due to the exclusion of galaxies with angular sizes that do not fit in the FoV of the integral field units (IFUs). These problems can be circumvented with multi-filter photometric surveys, that use a set of intermediate and narrow band filters designed to provide the required spectral information while still covering a large contiguous area.

\begin{figure}
\resizebox{\hsize}{!}{\includegraphics{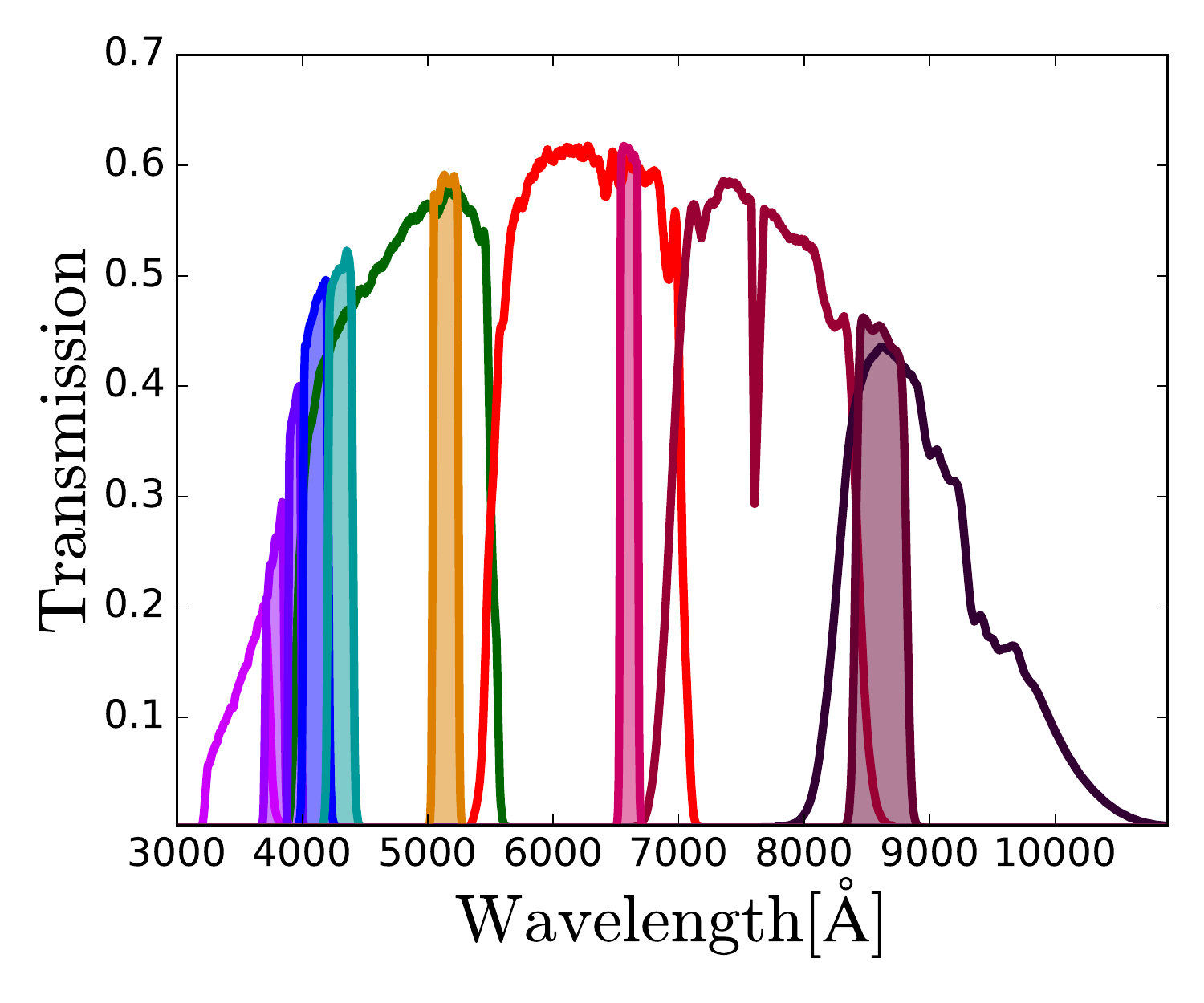}}
\caption{Transmission curves of the J-PLUS filter system; with the four broad ($g, r, i, z$), the two medium ($u, J0861$), and the six narrow ($J0378, J0395, J0410, J0430, J0515, J0660$) band filters. The transmission curves are computed after accounting for the effects caused by the efficiency of the CCD and the atmospheric extinction.}
\label{figure1}
\end{figure}

The Javalambre Photometric Local Universe Survey (J-PLUS\footnote{www.j-plus.es}, \citealt{cenarro18}) is currently operating to observe thousands of square degrees of the northern sky from the Observatorio Astrofísico de Javalambre 
(OAJ\footnote{http://oajweb.cefca.es/}) in Teruel, Spain. The survey is being carried out with the 0.83 meter JAST/T80 telescope and the panoramic camera T80Cam \citep{Toni}, with a 2 deg$^2$ FoV. A set of twelve broad, intermediate, and narrow band optical filters 
(Fig.~\ref{figure1} and Table~\ref{table1}) optimized to provide an adequate sampling of the spectral energy distribution (SED) of millions of stars in our galaxy is used.
These SEDs will be required for the photometric calibration of the Javalambre Physics of the accelerating universe Astrophysical Survey (J-PAS\footnote{www.j-pas.org}, \citealt{Benitez}). In addition, the position 
of the filters, the exposure times, and the survey strategy, are suitable to perform science that will expand our knowledge in many fields of astrophysics. Further details on the OAJ, the instrumentation, the filter set, the J-PLUS photometric calibration process, 
the strategy, and several science applications can be found in the J-PLUS presentation paper \citep{cenarro18}.

\begin{table*}
\caption{J-PLUS filters characteristics and limiting magnitudes} 
\label{table1} 
\centering 
\begin{tabular}{c c c c c c c c}
\hline \hline
Filter & Central & Pivot  & FWHM ($\rm nm$) & $\rm m_{\rm lim}$\tablefootmark{a} (SVD 1500041) & $\rm m_{\rm lim}$\tablefootmark{a}  (EDR) & $\sigma_{ZP}$&Spectral \\ 
 & wavelength ($\rm nm$) & wavelength ($\rm nm$) & & & & & Feature \\
\hline
 $u$&348.5    & 352.3  &50.8  &22.5  &21.5& 0.05  & -- \\ 
 $J0378$&378.5& 378.6  &16.8  &21.9  &21.4& 0.06 &[OII]   \\ 
 $J0395$&395.0& 395.1  &10.0  &21.3  &21.3& 0.07  &CaH+K  \\ 
 $J0410$&410.0& 410.1  &20.0  &21.0 &21.4& 0.06 &H$_{\rm \delta}$  \\ 
 $J0430$&430.0& 430.0  &20.0  &21.3  &21.4& 0.06 &$G$-Band\\ 
 $g$&480.3    & 474.5  &140.9  &23.1  &22.1& 0.05 & --  \\ 
 $J0515$&515.0& 515.0  &20.0  &21.4  &21.3& 0.04 &Mg$b$ Triplet  \\ 
 $r$&625.4    & 623.0  &138.8  &23.1  &21.8& 0.04 & -- \\ 
 $J0660$&660.0& 660.0  &13.8  &22.5  &21.1& 0.04 &${\rm H}\alpha$+[NII]\\ 
 $i$&766.8    & 767.7  &153.5  &22.0  &20.7& 0.04 & --  \\ 
 $J0861$&861.0& 860.3  &40.0  &22.1  &20.6& 0.05 &Ca Triplet  \\ 
 $z$&911.4    & 892.2  &140.9  &21.2  &20.5& 0.02 & --  \\  
\hline 
\end{tabular}
\tablefoot{\\
\tablefoottext{a}{The limiting magnitudes were measured at a fixed 3'' aperture for a signal to noise ratio ($\rm S/N$) of 3.}
}
\end{table*}

\begin{figure*}
\resizebox{\hsize}{!}{\includegraphics{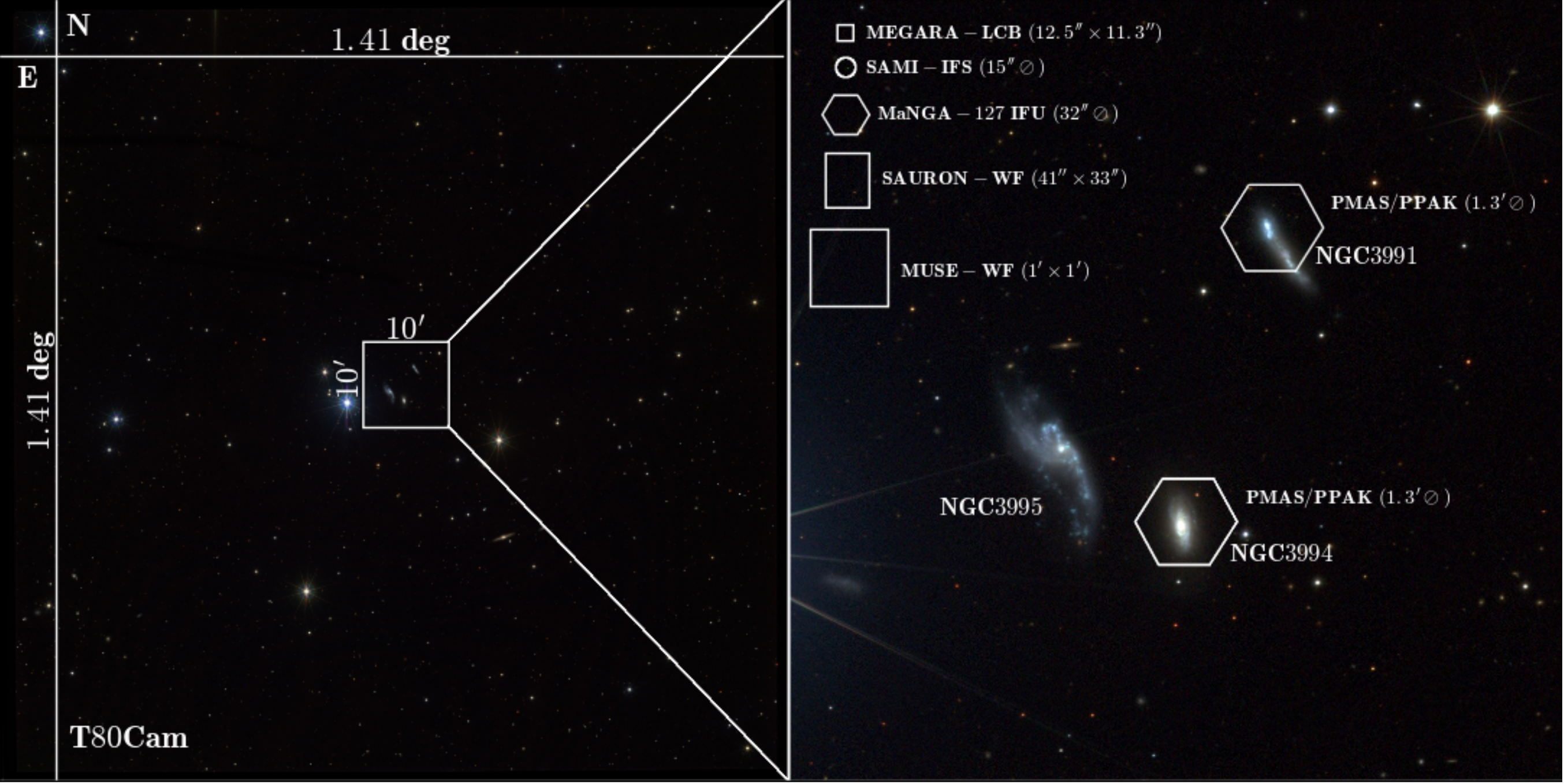}}
\caption{{\it Left panel}: colour composite image of the SVD pointing "1500041-Arp313", illustrating the 2deg$^{2}$ FoV of T80Cam at JAST/T80. {\it Right panel}:  $10^{\rm \prime}\times 10^{\rm \prime}$ zoom covering the Arp313 triplet, where the galaxies NGC3991, NGC3994, and NGC3995 are visible. The FoV of several IFUs is displayed: MEGARA \citep{megara}, SAMI \citep{Croom}, MaNGA \citep{Bundy,mangaifu}, SAURON \citep{SAURON}, MUSE \citep{MUSE} and PMAS/PPAK \citep{PPAK,CALIFA}. NGC3991 and NGC3994 are part of the CALIFA DR3 sample \citep{califadr3}.}
\label{figure8}
\end{figure*}

Among all the possible applications, we focus on star formation in the nearby universe $(\rm z \le 0.015)$, where the ${\rm H}\alpha$ emission line flux is covered by the $J0660$ narrow-band filter. J-PLUS offers a large contiguous area covered with twelve optical bands, enabling us to: (1) perform studies over the whole extent of galaxies, with no aperture selection effects; (2) carry out environmental studies; (3) count on a large statistical sample. These advantages are illustrated by showing the FoV of the T80Cam at JAST/T80 in Figure~\ref{figure8}, compared to the FoVs of several IFUs. The main star formation studies that can be addressed with the J-PLUS dataset are:

\begin{itemize}
 \item \textit{2D star formation properties:} benefiting from the characteristics of J-PLUS, it is possible to look into the spatially resolved star formation properties of nearby galaxies. With a large sample, it will be possible to study the correlation of these properties with morphology \citep[e.g. CALIFA:][]{Rosa}, the presence of close companions \citep[e.g. CALIFA:][]{barrera, cortijo}, environmental density \citep[e.g. SAMI:][]{schaefer}, or nuclear activity \citep[e.g. MaNGA:][]{sanchezAGN}, among others;
 \item \textit{HII region statistical studies:} it is possible to retrieve the excess of ${\rm H}\alpha$ flux over the underlying continuum in every pixel to construct HII region maps for spatially resolved galaxies. With these maps, it will be able to study the properties of HII regions in individual galaxies \citep[e.g. CALIFA:][]{sanchezhii};
 \item \textit{SFR density in the nearby universe:} the ${\rm H}\alpha$ luminosity function in the nearby universe can be computed, and the SFR density at $\rm z \le 0.015$ calculated. Similar published studies are based on a selected sample of galaxies \citep[e.g.][]{gallego, bothwell, Rosa}, but the depth and coverage of the J-PLUS observations allows the construction of a non pre-selected catalog of ${\rm H}\alpha$ emitters.
\end{itemize}

These topics will be presented in future publications, when the area surveyed by J-PLUS ensures a statistically meaningful sample. Once J-PLUS has finalized its observations, it is expected to have data for approximately 5,000 galaxies at $\rm z \le 0.015$. This represents a significant increase with respect to IFS surveys such as MaNGA, CALIFA, SAMI or VENGA with 524, 251, 127 and 30 galaxies at the same redshift range, respectively \citep{wake,califadr3,Croom,VIRUS}.

To carry out the above science cases, a reliable measurement of the ${\rm H}\alpha$ flux from J-PLUS data is required. In a previous work, \citet{Gonzalo15} predict, using synthetic photometry, that unbiased ${\rm H}\alpha$ fluxes from star-forming regions can be retrieved using the J-PLUS photometric system. The goal of the present paper is to test and validate the methodology presented by \citet{Gonzalo15} with real data, comparing the J-PLUS measurements with spectroscopic ${\rm H}\alpha$ emission line fluxes from the literature. SDSS and CALIFA were chosen so the comparison is performed with both spectroscopic fiber and integral field spectroscopic data, and because both datasets contain galaxies at $\rm z \le 0.015$. As of the date of this study, MaNGA currently released data did not contain any of the galaxies of the present study in this redshift range. The validation of the methodology will allow the study of the SFR in the neaby universe using the J-PLUS survey, and its southern counterpart, the Southern-Photometric Local Universe Survey (S-PLUS; \citealt{Claudia}). S-PLUS will observe thousands of square degrees of the southern sky with an identical telescope, camera and filter-set as J-PLUS, making possible the use of all the tools and methodologies developed for the northern survey.

This paper is organized as follows: we present the J-PLUS and spectroscopic data used in this work in Sect.~\ref{data}. Section~\ref{methodologies} is devoted to explain the calculation of the ${\rm H}\alpha$ emission line fluxes with the different datasets. In Sect.~\ref{results_section}, the comparison between J-PLUS ${\rm H}\alpha$ fluxes and the spectroscopic ones is presented. The SFR science that will be tackled with J-PLUS is exposed with an example of the spatially resolved SFR in a close pair of galaxies in Sect.~\ref{science}. Finally, we summarize the work done and present the conclusions in Sect.~\ref{summary}. A standard cosmology $H_{0}=70 \ \rm km \ s^{-1} \ Mpc^{-1}$, $\Omega_{\rm m}=0.3$, $\Omega_{\Lambda}=0.7$ and $\Omega_{\rm k}=0$ is used. Magnitudes are expressed in the AB system \citep{AB}. 
                                                                                                                                                                                                                                                                                                                                                                                                                                                                                                                                                                                                                              
\section{J-PLUS and spectroscopic data}\label{data}

In this section, the different datasets used for the work are presented. We first describe the photometric J-PLUS data acquisition and reduction processes, and give the main characteristics of the data (Sect.~\ref{J-PLUS-data}). Then, the spectroscopic data from SDSS (Sect.~\ref{23}) and the CALIFA survey (Sect.~\ref{CALIFA_data}) used in this comparison study are introduced. 

\subsection{J-PLUS photometric data}\label{J-PLUS-data}
The photometric data used in this work were acquired with the 0.83 meter JAST/T80 telescope and the panoramic camera T80Cam, set up with a 9216  \texttimes \ 9232 pixel CCD. The system offers
a 1.4 \ \texttimes \ 1.4 deg$^2$ FoV with a 0.55$^{''}$/pixel scale. Figure~\ref{figure1} shows the transmission curves of the complete set of filters, and Table~\ref{table1}
displays their characteristics. The fields for the study were chosen in order to have ${\rm H}\alpha$ measurements of star-forming regions in common with SDSS, and galaxies with star formation activity in common with CALIFA, both at $ \rm z \le 0.015$.
The final eight pointings selected belong to two different datasets. One is the J-PLUS Early Data Release (EDR), a subset of 36 deg$^{2}$ representative of J-PLUS data in terms of depth, point spread function (PSF), and photometric calibration accuracy \citep{cenarro18}.
The other is a J-PLUS Science Verification Data project (SVD 1500041; P.I.: G. Vilella-Rojo), that was especially planned for this work. The observations of the J-PLUS SVD 1500041 took place during several nights in February 2016 and
 on average, they are 0.85 magnitudes deeper than J-PLUS EDR. They were required to be observed with these depths for being part of the science verification of the T80Cam at JAST/T80 system. In the following, we refer to the joint SVD 1500041 and EDR data as ``J-PLUS'' data. Further details on the limiting magnitudes of both datasets and the characteristics of the selected fields can be found in Tables ~\ref{table1} and ~\ref{table2}, respectively.

 In addition to the present paper, the J-PLUS EDR and SVD were used to refine the membership in nearby galaxy clusters \citep{molino18}, analyse the globular cluster M15 \citep{bonatto18}, study the stellar populations of several local galaxies \citep{sanroman18}, and compute the stellar and galaxy number counts up to $r = 21$ \citep{clsj18}.

\begin{table*}
\caption{Description of the T80Cam pointings} 
\label{table2} 
\centering 
\begin{tabular}{c c c c c c}
\hline \hline
J-PLUS  &Central  &Central& SDSS   & CALIFA   & Comments \\
pointing&RA (º)   &DEC (º)    & Star forming   &  Star forming &   \\
&   &  & regions  &  galaxies &   \\
\hline
1500041-M101 &211.105  &54.650 &8  &---- &M101 is present\\ 
1500041-Arp313 &179.423  &32.298  &10  &NGC3991 \& NGC3994  &Arp 313 triplet is present\\ 
1500041-M49 &188.042 &8.134&3  &NGC4470  &M49 is present\\
JPLUS-00745&121.809&30.443&1&----&----\\
JPLUS-00749&128.213&30.443 &1&----&----\\
JPLUS-01500&131.116&40.190 &1&----&----\\
JPLUS-01506&141.923&40.190 &1&----&----\\
JPLUS-01588&109.596&41.582 &1&----&----\\
\hline 
\end{tabular}
\end{table*}

Once the pointings were observed; data reduction, photometric calibration, and image coadding were automatically performed by different pipelines included in the \texttt{Jype} package, fully developed by the Data Processing and Archiving Unit (UPAD) team at CEFCA. The final products for every pointing are photometric catalogs of the sources found, as well as coadded images and weight maps in the twelve bands with calibration zero-points ($ZPs$). The uncertainties in the calibration ($\sigma_{ZP}$) can be found in Table~\ref{table1}. Further details about the processes involved in the production of the images and catalogs are explained in \citet{cenarro18}. In this study the coadded images were used, but before working with them, extra operations were needed in addition to the standard automatic procedures. 

A PSF homogenization in the twelve bands for each pointing was performed. The homogenization of the PSF is required to provide good quality photometry, because the coadded images are constructed with different individual ones. This means that several exposures are combined to create a single final image, each of them not necesarily having the same PSF, generating inhomogeneities in the light distribution of the sources in a given band. Additionally, the PSF associated with a given object depends on the band. As a result of this, the light inside a given aperture on the object is also distributed differently depending on the band. 

\begin{figure}
\resizebox{\hsize}{!}{\includegraphics{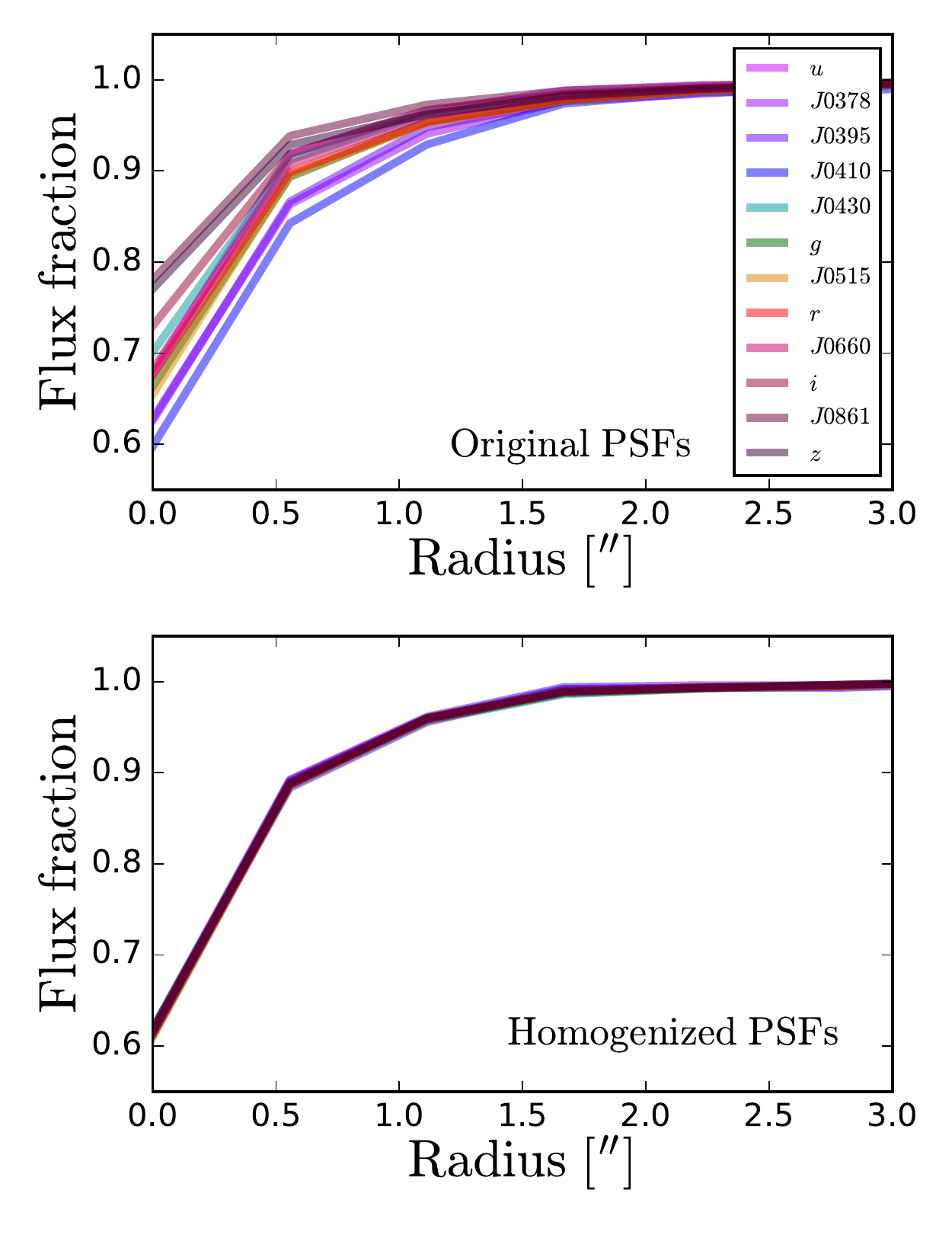}}
\caption{Effect of the PSF homogenization on the light profile of an example source. {\it Top:}  the lines with different colors correspond to the light profile of the source in the twelve J-PLUS bands. {\it Bottom:} the light is equally distributed in every band after the homogenization, and the profiles are indistinguishible.}
\label{figure2}
\end{figure}

These effects on the PSF may produce artificial structures that could bias the results \citep{Bertin1}. To deal with them, we used the \texttt{JypePSF} module inside the \texttt{Jype} package. The \texttt{JypePSF} module analyzes, characterizes, and homogenizes the PSF in the following way:
\begin{itemize}
 \item The worst PSF value of each pointing is computed by checking the full width at half maximum of the PSF ($\rm PSF_{\rm FWHM}$) of bright point sources in every band;
 \item The obtained value is selected as the target for the homogenization in the twelve bands; although the code also allows the user to choose the reference value;
 \item A homogenization kernel is created for different positions in every image. In this step, \texttt{SExtractor} \citep{Bertin3} and \texttt{PSFex} \citep{Bertin2} are used;
 \item The images are convolved with their corresponding kernels using a fast Fourier transform,
 bringing the images of all the bands to the same circular PSF;
 \item This process has consequences in the image noise, introducing correlations that need to be considered. Therefore, the background model of the images is recalculated, since it is later used to 
 compute photometric errors (Sect.~\ref{eqs}).
\end{itemize}
The correct performance of the PSF homogenization process can be checked by looking at the light profile of the sources, as the example in Fig.~\ref{figure2}.

\begin{figure*}
\resizebox{\hsize}{!}{\includegraphics{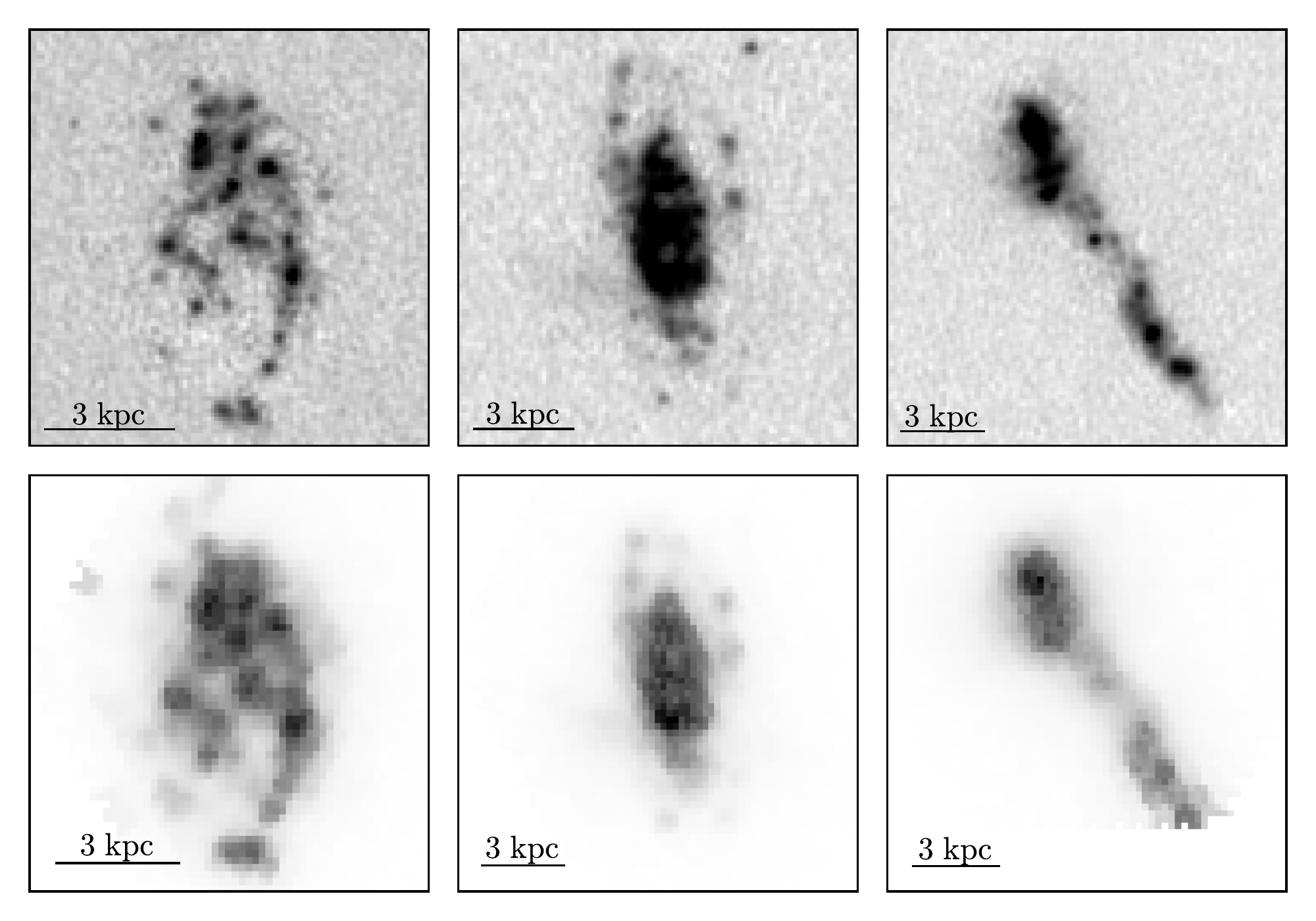}}
\caption{J-PLUS images showing ${\rm H}\alpha$ emission areas with the excess in the J0660 band, $J0660 \ - \ r$ ({\it top panels}), and CALIFA ${\rm H}\alpha$ maps ({\it bottom panels}) of the selected star-forming galaxies: NGC4470 ({\it left panels}), NGC3994 ({\it center panels}), and NGC3991 ({\it right panels}). For reference, north is up and east is left.}
\label{figure4}
\end{figure*}

Two sets of J-PLUS images were prepared for every pointing, in which the PSF homogenization was performed by choosing the target value according to the median $\rm PSF_{\rm FWHM}$ of SDSS-DR12 and CALIFA-DR2, respectively. The median $\rm PSF_{\rm FWHM}$ of SDSS-DR12 is 1.30$''$ \citep{dr12}, similar to the one of the J-PLUS sample. In the case of CALIFA-DR2 the value is 2.4$''$ \citep{GarciaBenito}, due to the different spatial sampling, rather than to observing conditions (Fig.~\ref{figure4}). 

\subsection{SDSS data}\label{23}
The SDSS data used in this work, consist of emission line flux measurements of star-forming regions made by the Portsmouth group \citep{tomas} and their corresponding original spectra. They are part of the SDSS 12nd Data Release (DR12, \citealt{dr12}; including BOSS). These measurements are based on observations of nearby galaxies taken with multiple fiber spectra, as a result of a wrong deblending due to the apparent large size of these objects. They were collected with the Sloan foundation 2.5m telescope at Apache Point observatory. The complete set of data is available for download in the DR12 website\footnote{https://dr12.sdss.org/advancedSearch}$^{,}$\footnote{http://www.sdss.org/dr12/spectro/galaxy\_portsmouth}. The emission line measurements by \citet{tomas} are performed with an adaptation of the publicly available code Gas and Absorption Line Fitting code (\texttt{GANDALF v1.5}, \citealt{sarzi}), by fitting stellar population models and gaussian emission line templates to the spectra. The stellar population models from \citet{maraston}, based on the MILES stellar library, are used in this task.

\subsection{CALIFA data}\label{CALIFA_data}
The CALIFA data used for this work belong to the 2nd Data Release \citep{CALIFA,sample,GarciaBenito}, based on observations collected at the Centro Astronómico Hispano Alemán (CAHA) at Calar Alto with the PMAS/PPAK integral field spectrophotometer, mounted on the Calar Alto 3.5 m telescope. Data products can be found in the CALIFA website\footnote{www.caha.es/CALIFA/}, from where we downloaded the V500 data cubes
for the galaxies included in the present study. The three CALIFA galaxies selected are star-forming spirals at $\rm z \le 0.015$ (Table~\ref{table2}).

In addition to the data cubes, ${\rm H}\alpha$ measurements from CALIFA galaxies are required to
do the comparison. We used the available data products provided by \texttt{Pipe3D} in the CALIFA-DR2 website\footnote{ftp://ftp.caha.es/CALIFA/dataproducts/DR2/Pipe3D/}, where ${\rm H}\alpha$ emission maps of the galaxies in the present research are included (Fig.~\ref{figure4}).

\texttt{Pipe3D} is a pipeline based on the \texttt{FIT3D} fitting tool \citep{fit3D1,fit3D2}. It is designed to study the properties of the stellar populations and ionized gas of IFS data, thus being a higlhy suitable tool for the analysis of the data cubes from ongoing and upcoming IFU surveys. Among the several steps performed by the pipeline, we higlhight: (1) the stellar continuum fitting with stellar population models by the MILES project \citep{Patricia,vazdekis,falcon}. This is done on binned spaxels of the data cubes, according to S/N criteria; (2) the decoupling of every spaxel along with their own spectra in the bins; (3) the fitting of the strongest spectral emission lines for every spaxel, and production of spatially resolved maps of the flux intensity of the lines.
Detailed information about \texttt{Pipe3D} can be found in \citet{Sanchez2,Sanchez1}.

\section{Measuring ${\rm H}\alpha$ in the nearby universe}\label{methodologies}

In this section, the steps to obtain the ${\rm H}\alpha$ flux measurements from spectroscopic data (Sect.~\ref{flux_from_SDSS} and ~\ref{bptdiag}) and from J-PLUS photometric data (Sect.~\ref{eqs}) are explained. The final goal is to have the emission line fluxes from the different datasets measured in the same regions, making the comparison presented in Sect. ~\ref{results_section} possible.

\subsection{Flux measurements from SDSS}\label{flux_from_SDSS}
 We initially selected all the sources in the J-PLUS fields with emission line measurements taken by the Portsmouth group from SDSS data. Then, a series of selection criteria were applied to compose the final sample:
 \begin{itemize}
  \item The Portsmouth group includes for every source a classification according to a Baldwin, Phillips \& Terlevich (BPT) diagram \citep{BPT}; we used this classification to choose sources that are classified as 'Star Forming / HII', avoiding AGN contamination in the sample. This step is necessary, since SDSS fibers are usually located in the center of galaxies, where AGN effects are expected to be most important;
  \item Only star-forming regions that are at $\rm z \le 0.015$ remained in the sample, since these are representative of the ones we deal with in J-PLUS;
  \item Considering the equivalent width measurements of the lines by the Portsmouth group, we imposed a lower limit of 12 $\AA$ in the ${\rm H}\alpha$ equivalent width ($\rm EW_{{\rm H}\alpha} \ge 12 \AA$). This is done beacuse J-PLUS cannot resolve spectroscopically the ${\rm H}\alpha$ emission line with a 3$\sigma$ precision if it has an equivalent width below that value (eq. [13] in \citealt{Gonzalo15}).   
 \end{itemize}
   The final sample contains 26 star-forming regions. The ${\rm H}\alpha$ fluxes and errors of the regions provided by the Portsmouth group are dust-corrected with \citet{calzetti} dust extinction law. The numerical values are reported in Table~\ref{table3}.
 
\subsection{Flux measurements from CALIFA}\label{bptdiag}

\begin{figure}
\resizebox{\hsize}{!}{\includegraphics{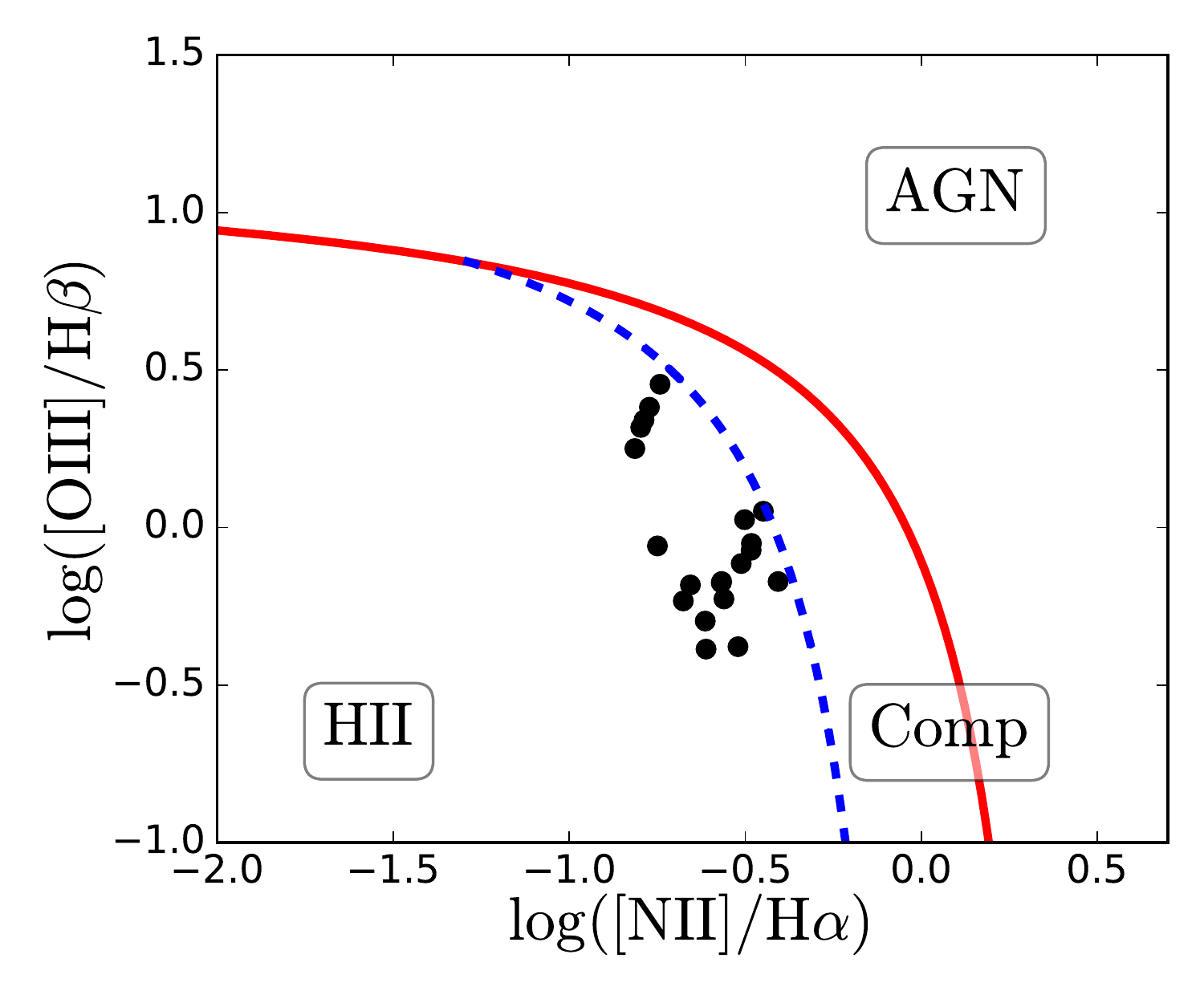}}
\caption{BPT diagram based on CALIFA data of the 20 selected star-forming regions in NGC3991, NGC3994 and NGC4470. The blue-dashed and red lines delimitate the composite and AGN zone, respectively \citep{kewley,kauffmann}.}
\label{figure3}
\end{figure}

 The \texttt{Pipe3D} ${\rm H}\alpha$ flux maps derived from CALIFA data \citep{Sanchez2,Sanchez1} were used to visually select star-forming region candidates. We placed 3$''$ diameter circular apertures, mimicking SDSS fibers, on the knots of star formation, then we imposed them to fulfill the same criteria required for SDSS star-forming regions (Sect.~\ref{flux_from_SDSS}). The selected candidates are all at $\rm z \le 0.015$ with $\rm EW_{{\rm H}\alpha} \ge 12 \AA$. To compute their BPT diagram, the different fluxes from the different \texttt{Pipe3D} emission line flux maps available were extracted \citep{Sanchez2,Sanchez1}. This task was carried out with the \texttt{funcnts} module included in the \texttt{funtools} package \citep{funtools}, that allows to perform aperture photometry measurements on selected regions. Once the emission lines were extracted, the regions were placed in the BPT diagram (Fig.~\ref{figure3}). The diagram is divided in HII, composite, and AGN zones \citep{kewley,kauffmann}. The 20 candidates to star-forming regions are within the HII area, accomplishing the established criteria. Eight regions are included in both NGC3994 and NGC4470, and four are included in NGC3991 (Table~\ref{table3} and Fig.~\ref{figure4}). 

\begin{table*}
\caption{Selected star-forming regions for the comparison between the spectroscopic and J-PLUS $\rm H\alpha$ emission line fluxes} 
\label{table3} 
\centering 
\begin{tabular}{c c c c c}
\hline \hline
Region &RA (º)&DEC (º) & $\log F^{\rm J-PLUS}_{\rm H\alpha}$  & $\log F^{\rm spec}_{\rm H\alpha}$  \\
       &      &        & ${\rm[10^{-17} erg \ s^{-1} \ cm^{-2}]}$  &        ${\rm[10^{-17} erg \ s^{-1} \ cm^{-2}]}$          \\                                                
\hline             
CALIFA-NGC4470-1 & 187.409 & 7.826       & 3.41 $\pm$ 0.07  & 3.39 $\pm$  0.04 \\
CALIFA-NGC4470-2 & 187.407 & 7.827       & 3.28 $\pm$ 0.07  & 3.34 $\pm$  0.02 \\
CALIFA-NGC4470-3 & 187.406 & 7.826       & 3.37 $\pm$ 0.07  & 3.35 $\pm$  0.03 \\
CALIFA-NGC4470-4 & 187.408 & 7.825       & 3.34 $\pm$ 0.08  & 3.44 $\pm$  0.04 \\
CALIFA-NGC4470-5 & 187.407 & 7.823       & 3.20 $\pm$ 0.09  & 3.25 $\pm$  0.03 \\
CALIFA-NGC4470-6 & 187.410 & 7.823       & 3.27 $\pm$ 0.08  & 3.16 $\pm$  0.04 \\
CALIFA-NGC4470-7 & 187.405 & 7.823       & 3.15 $\pm$ 0.08  & 3.22 $\pm$  0.03 \\
CALIFA-NGC4470-8 & 187.405 & 7.822       & 3.37 $\pm$ 0.07  & 3.50 $\pm$  0.03 \\
CALIFA-NGC3994-1 & 179.405 & 32.283      & 2.81 $\pm$ 0.07  & 2.77 $\pm$  0.01 \\
CALIFA-NGC3994-2 & 179.400 & 32.280      & 2.90 $\pm$ 0.07  & 2.78 $\pm$  0.01 \\
CALIFA-NGC3994-3 & 179.406 & 32.281      & 2.92 $\pm$ 0.07  & 2.90 $\pm$  0.01 \\
CALIFA-NGC3994-4 & 179.404 & 32.271      & 2.21 $\pm$ 0.09  & 2.24 $\pm$  0.01 \\
CALIFA-NGC3994-5 & 179.400 & 32.278      & 2.85 $\pm$ 0.07  & 2.64 $\pm$  0.01 \\
CALIFA-NGC3994-6 & 179.402 & 32.273      & 2.80 $\pm$ 0.08  & 2.74 $\pm$  0.01 \\
CALIFA-NGC3994-7 & 179.403 & 32.276      & 3.86 $\pm$ 0.07  & 4.35 $\pm$  0.06 \\
CALIFA-NGC3994-8 & 179.403 & 32.277      & 4.09 $\pm$ 0.11  & 3.88 $\pm$  0.04 \\
CALIFA-NGC3991-1 & 179.375 & 32.333      & 3.44 $\pm$ 0.07  & 3.78 $\pm$  0.04 \\
CALIFA-NGC3991-2 & 179.381 & 32.339      & 3.52 $\pm$ 0.07  & 3.87 $\pm$  0.03 \\
CALIFA-NGC3991-3 & 179.381 & 32.339      & 3.50 $\pm$ 0.07  & 3.73 $\pm$  0.02 \\
CALIFA-NGC3991-4 & 179.382 & 32.341      & 3.94 $\pm$ 0.07  & 4.20 $\pm$  0.04 \\
SDSS-1627-53473-0554 & 188.194 & 7.799   & 2.07 $\pm$ 0.12  & 2.09 $\pm$  0.02 \\
SDSS-1627-53473-0611 & 188.576 & 8.239   & 3.64 $\pm$ 0.07  & 3.61 $\pm$  0.02 \\
SDSS-1628-53474-0323 & 188.589 & 8.240   & 3.16 $\pm$ 0.07  & 3.11 $\pm$  0.04 \\
SDSS-1323-52797-0108 & 209.971 & 54.786  & 3.11 $\pm$ 0.07  & 3.11 $\pm$  0.01 \\
SDSS-1324-53088-0183 & 211.388 & 54.461  & 3.74 $\pm$ 0.07  & 3.58 $\pm$  0.01 \\
SDSS-1324-53088-0386 & 210.135 & 55.246  & 2.79 $\pm$ 0.09  & 2.87 $\pm$  0.01 \\
SDSS-1325-52762-0398 & 211.341 & 54.269  & 2.96 $\pm$ 0.10  & 2.78 $\pm$  0.04 \\
BOSS-6739-56393-0662 & 210.827 & 54.589  & 1.84 $\pm$ 0.11  & 1.91 $\pm$  0.01 \\
BOSS-6797-56426-0464 & 211.384 & 54.459  & 2.13 $\pm$ 0.11  & 2.10 $\pm$  0.01 \\
BOSS-6797-56426-0488 & 210.919 & 54.496  & 1.80 $\pm$ 0.15  & 1.33 $\pm$  0.04 \\
BOSS-6801-56487-0008 & 212.111 & 54.908  & 2.35 $\pm$ 0.10  & 2.38 $\pm$  0.03 \\
SDSS-1991-53446-0553 & 178.927 & 32.188  & 3.32 $\pm$ 0.11  & 3.06 $\pm$  0.05 \\
SDSS-1991-53446-0571 & 179.168 & 32.628  & 2.56 $\pm$ 0.13  & 2.46 $\pm$  0.01 \\
SDSS-1991-53446-0587 & 179.435 & 32.298  & 3.53 $\pm$ 0.08  & 3.48 $\pm$  0.02 \\
SDSS-1991-53446-0588 & 179.427 & 32.284  & 3.42 $\pm$ 0.07  & 3.43 $\pm$  0.01 \\
SDSS-1991-53446-0592 & 179.135 & 32.130  & 2.33 $\pm$ 0.11  & 2.22 $\pm$  0.01 \\
SDSS-1991-53446-0600 & 179.096 & 32.038  & 2.54 $\pm$ 0.15  & 2.36 $\pm$  0.04 \\
SDSS-2095-53474-0311 & 179.142 & 32.128  & 2.38 $\pm$ 0.10  & 2.35 $\pm$  0.01 \\
SDSS-2095-53474-0352 & 179.381 & 32.339  & 3.79 $\pm$ 0.07  & 3.81 $\pm$  0.02 \\
SDSS-2095-53474-0355 & 179.433 & 32.294  & 4.02 $\pm$ 0.07  & 4.07 $\pm$  0.03 \\
BOSS-4601-55589-0198 & 179.427 & 32.285  & 2.70 $\pm$ 0.08  & 2.62 $\pm$  0.02 \\
SDSS-0860-52319-0468 & 121.067 & 30.182  & 2.92 $\pm$ 0.08  & 2.83 $\pm$  0.04 \\
SDSS-1208-52672-0399 & 127.416 & 31.077  & 3.44 $\pm$ 0.07  & 3.57 $\pm$  0.01 \\
SDSS-0829-52296-0434 & 130.908 & 40.429  & 2.67 $\pm$ 0.09  & 2.60 $\pm$  0.01 \\
SDSS-0939-52636-0081 & 142.065 & 40.149  & 2.75 $\pm$ 0.09  & 2.66 $\pm$  0.01 \\
SDSS-1864-53313-0249 & 110.280 & 41.127  & 2.77 $\pm$ 0.08  & 2.78 $\pm$  0.01 \\
\hline 
\end{tabular}
\end{table*}

The ${\rm H}\alpha$ emission line fluxes of the CALIFA selected star-forming regions need to be corrected for dust extinction. To do so, we repeated the procedure applied to the SDSS data by using the \citet{calzetti} extinction law with the Balmer decrement. The following expressions need to be applied, 
\begin{equation}
 F_{\rm i} = F_{\rm o}10^{0.4E(B-V)k'(\lambda)} = F_{\rm o}10^{1.33E(B-V)},
\end{equation}
where $F_{\rm i}$ and $F_{\rm o}$ represent the intrinsic and observed flux respectively, and $k'(\lambda) = 3.33$ for ${\rm H}\alpha$. The colour excess $E(B-V)$ is 
\begin{equation}
 E(B-V)=1.97\log_{10}\Bigg[ \frac{({\rm H}\alpha / {\rm H}\beta)_{\rm o}}{2.86}\Bigg].
\end{equation}

The fluxes and their errors are retrieved from \texttt{Pipe3D} emission line flux maps. The errors of the fluxes in a given aperture need to be calculated following the prescriptions in \citet{GarciaBenito}. Numerical values of the ${\rm H}\alpha$ emission line fluxes of the CALIFA star-forming regions are reported in Table~\ref{table3}.

\subsection{Flux meauserements from J-PLUS data}\label{eqs}

In order to measure the ${\rm H}\alpha$ emission line fluxes of the SDSS and CALIFA star-forming regions in J-PLUS data, we started by retrieving a catalog of them, with instrumental photometry in the twelve bands. This was done performing aperture photometry on the same spectroscopic locations (Table~\ref{table3}) and mimicking the same apertures, with the \texttt{funcnts} software. The instrumental photometry is measured in counts ($C$), expressed in analog to digital units ($\rm ADUs$). The flux density of a source in a given filter is 
\begin{equation}
F_{\rm \lambda}=(C-C_{\rm B}) \ 10^{-0.4(ZP+48.6)}\frac{c}{\lambda^{2}_{\rm pivot}},
\end{equation}
where $C_{\rm B}$ represents the background counts, $ZP$ is the calibration zero point of the band, $c$ is the speed of light, and $\lambda_{\rm pivot}$ is the pivot wavelength of the filter, which is a source-independent measure of the characteristic wavelength of a bandpass given by 
\begin{equation}
\lambda^{2}_{\rm pivot} = \frac{\int T(\lambda)d\lambda}{\int{\frac{T(\lambda)}{\lambda^2}}d\lambda}, 
\end{equation}
where $T(\lambda)$ represents the transmission curve of the filter. The $\lambda_{\rm pivot}$ values of the J-PLUS filters are shown in Table~\ref{table1}. To calculate the errors, we have to take into account that there are three sources of uncertainty that affect the measurements: the uncertainty in the zero point ($\sigma_{ZP}$), the large scale background variation ($\sigma_{C_{\rm B}}$, \citealt{labbe,molino}), and the electron counting by the CCD ($\sigma_{C}$). The total uncertainty in the measurements for a source in a certain band is then given by
\begin{equation}
\sigma_{F}=\sqrt{\bigg(\frac{\partial{F_{\rm \lambda}}}{\partial{ZP}}\sigma_{ZP}\bigg)^{2}+\bigg(\frac{\partial{F_{\rm \lambda}}}{\partial{C_{\rm B}}}\sigma_{C_{\rm B}}\bigg)^{2}+\bigg(\frac{\partial{F_{\rm \lambda}}}{\partial{C}}\sigma_{C}\bigg)^{2}},
\label{zp_eq}
\end{equation}
where 
\begin{equation}
\sigma_{C_{\rm B}}=S_{\rm fit}\sqrt{N_{\rm pix}}(a_{\rm fit}+b_{\rm fit}\sqrt{N_{\rm pix}})
\end{equation}
and
\begin{equation}
\sigma_{C}=\sqrt{\frac{C}{G}}
\end{equation}
are the uncertainties associated to the background variation and the electron counting, respectively, $G$ is the gain of the detector, $N_{\rm pix}$ is the size of the aperture in pixels, and $S_{\rm fit}$, $a_{\rm fit}$, $b_{\rm fit}$ are the resulting coefficients from the background model (see Sect.~\ref{J-PLUS-data} for details). Fluxes and errors were calculated following the previous expressions. 

Two effects need to be taken into account before getting the final photometric catalog of the star-forming regions: (1) the astrometry in the different datasets data may have offsets that could introduce mismatches between the location of the regions; (2) the calibration processes differ from one survey to another, bringing up the requirement of having a common scale among them. 

\begin{figure*}
\resizebox{\hsize}{!}{\includegraphics{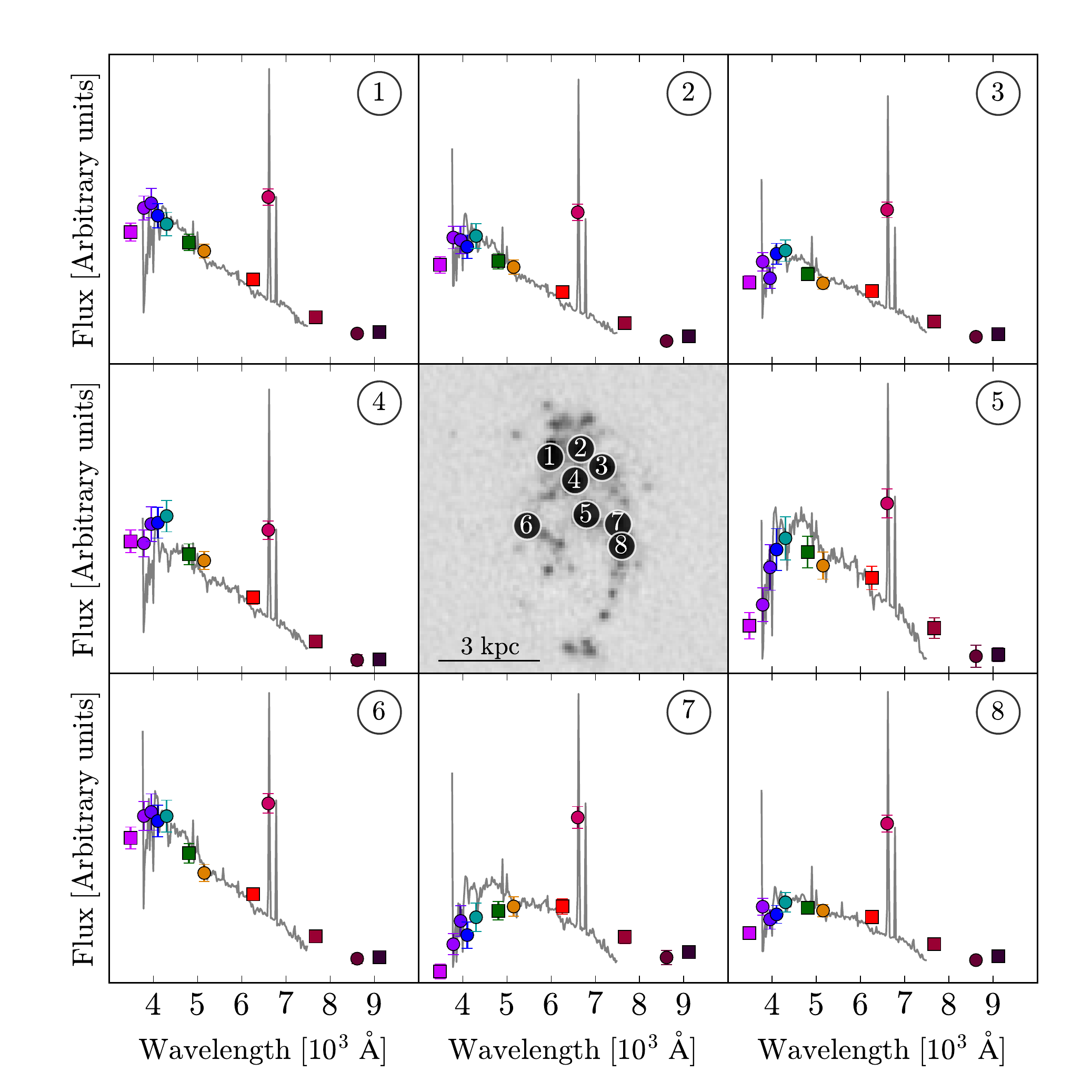}}
\caption{The eight selected star-forming regions in NGC4470. {\it Central panel}: J-PLUS image showing ${\rm H}\alpha$ emission areas with the excess in the J0660 band ($J0660 \ - \ r$), and the selected regions marked with numbers. {\it Surrounding panels} show J-PLUS SEDs of the corresponding regions, with the CALIFA spectra plotted in gray. The color code for the bands is the same than in Fig.~\ref{figure1} and Fig.~\ref{figure2}.}
\label{figure5}
\end{figure*}

In order to minimize aperture displacement effects, the offsets were inspected and corrected in the regions where it was required. To deal with calibration issues, synthetic photometry of the regions was created by convolving their full SDSS/CALIFA spectra with the J-PLUS filter set, then we matched the synthetic $r$-band fluxes with the measurement from J-PLUS images and scaled accordingly all the bands producing the final photometric catalog. The SEDs and spectra of the star-forming regions were plotted to check the agreement between the different datasets, finding an overall agreement that confirms the matching of the regions in terms of astrometry and calibration. As an example, NGC4470 selected star-forming regions from CALIFA data are presented in Fig.~\ref{figure5}.
 
 The following step consists in retrieving the dust corrected ${\rm H}\alpha$ emission line flux of every star-forming region. The methodology designed by \citet{Gonzalo15} to do it is briefly reviewed in the next section.
 
 \subsubsection{Extracting ${\rm H}\alpha$ emission line fluxes from J-PLUS photometric data}
 
 In a previous work, \citet{Gonzalo15} show that reliable ${\rm H}\alpha$ fluxes of star-forming regions in galaxies with $\rm z \le 0.015$ can be recovered  using the J-PLUS photometric data. In that redshift range, the ${\rm H}\alpha + \rm [NII]$ emission is covered by the $J0660$ narrow band filter, and different methodologies were tested to extract the ${\rm H}\alpha$ emission from it. Synthetic J-PLUS data from a sample of 7511 SDSS spectra with measured ${\rm H}\alpha$ flux were used for this task. The best results were achieved with a 3-step process that:
\begin{itemize}
 \item Removes the underlying stellar continuum and extracts the ${\rm H}\alpha + \rm [NII]$ emission, using a SED-fitting method based on simple stellar population template models from \citet{Bruzual}. The photometry in the twelve J-PLUS bands is required for this step;
\item Corrects the emission from dust extinction in the host galaxy, making use of a derived empirical relation based on the ($g$ - $i$) colour, obtained with SDSS data (eq. [20] in \citealt{Gonzalo15});
\item Removes the $\rm [NII]$ contribution and isolates the ${\rm H}\alpha$  emission line
flux with another derived empirical relation based on the ($g$ - $i$) colour as well (eq. [21] in \citealt{Gonzalo15}).
\end{itemize}
 The comparison between the ${\rm H}\alpha$ values derived from the synthetic photometry and the original spectroscopic ${\rm H}\alpha$ measurements proves that it is possible to get unbiased ${\rm H}\alpha$ emission line fluxes with J-PLUS in a robust and homogeneous way. All the details of the process can be found in \citet{Gonzalo15}. 

The dust corrected ${\rm H}\alpha$ fluxes from J-PLUS photometric data used in the present work were computed with this methodology, using the photometric catalogue with the SED of every star-forming region. The results are reported in Table~\ref{table3}.

\section{Results}\label{results_section}

\begin{figure*}
\resizebox{\hsize}{!}{\includegraphics{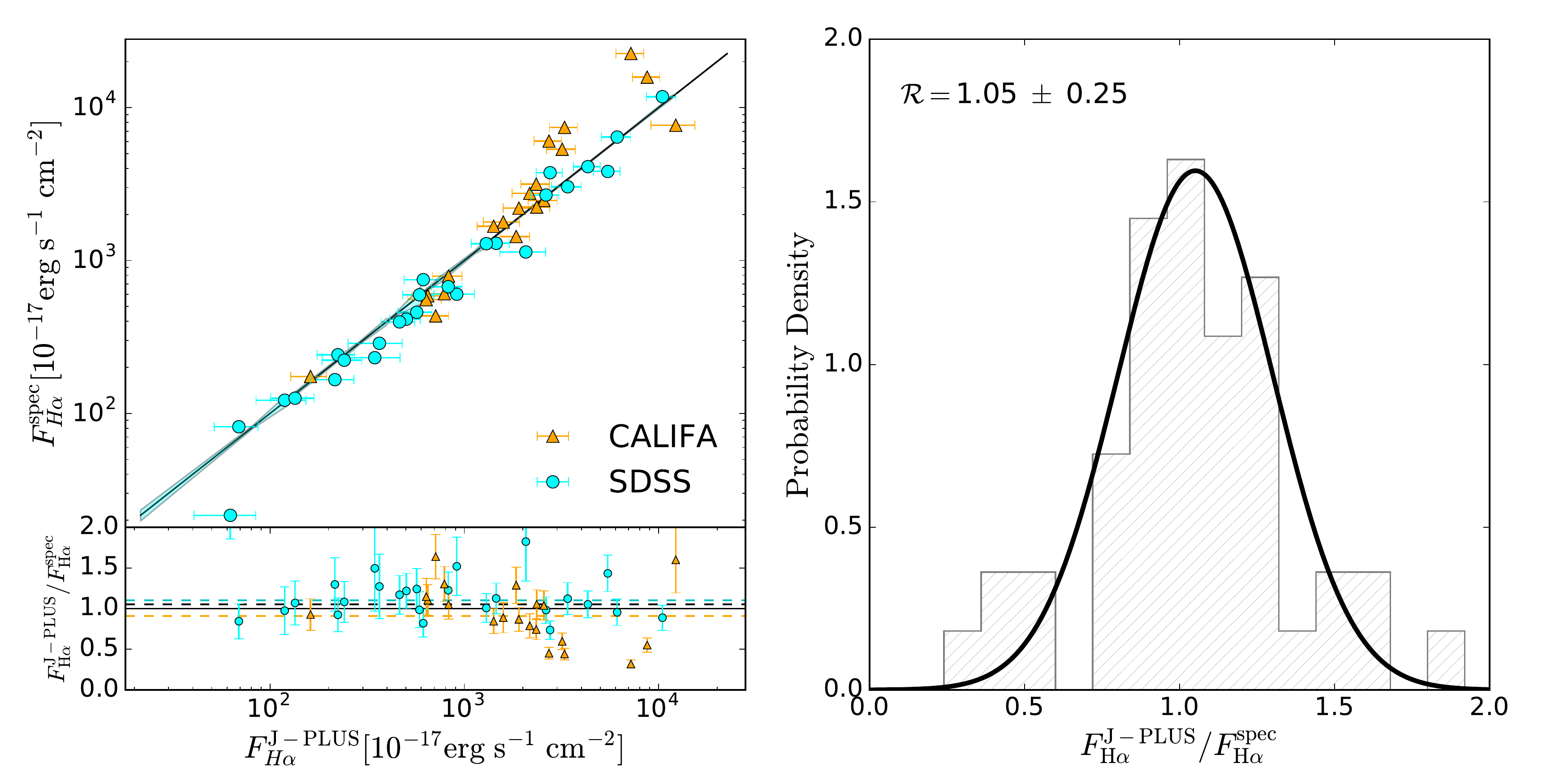}}
\caption{{\it Left panel}: comparison between J-PLUS and spectroscopic emission line fluxes. The errorbars represent the J-PLUS uncertainties and the shaded areas the spectroscopic ones. The ratio between the fluxes are in the lower panel, with the median ratios of CALIFA, SDSS and the total sample represented in orange, blue and black dashed lines, respectively. {\it Right panel}: normalised distribution of the flux ratios with a gaussian curve fitted. The median ratio $\mathcal{R}$ and its dispersion are displayed in the panel.}
\label{figure6}
\end{figure*}

In the previous section we have compiled the ${\rm H}\alpha$ emission line fluxes of 46 star-forming regions with both spectroscopic CALIFA and SDSS data and photometric J-PLUS data. Now the measurements of the emission line fluxes are compared by computing the ratio $\mathcal{R}$, defined as
 \begin{equation}\label{ratio}
  \mathcal{R}=F^{\rm J-PLUS}_{\rm H\alpha}/F^{\rm spec}_{\rm H\alpha},
 \end{equation}
 where $F^{\rm J-PLUS}_{\rm H\alpha}$ represents the J-PLUS photometric fluxes, and $F^{\rm spec}_{\rm H\alpha}$ represents the spectroscopic ones. The median ratio of the 46 star-forming regions analysed here yields $\mathcal{R} = 1.05$, with a $1\rm \sigma$ dispersion of $0.25$. The results are summarized in Fig.~\ref{figure6}.
 
 The comparison between the J-PLUS values and the spectroscopic measurements is consistent, being the median $\mathcal{R}$ close to the unit, with a small overestimation of 5\%. This deviation is not significant if we take into account the low number of regions in the study, and the fact that it lies far below the $1\rm \sigma$ uncertainty given by the dispersion. This result is close to the expectations by \citet{Gonzalo15}, that find a median comparison ratio of $\mathcal{R} = 0.99 \pm 0.15$ computed with synthetic J-PLUS photometry. This demonstrates that unbiased ${\rm H}\alpha$ emission line fluxes can be measured using J-PLUS photometry.

However, we find that only 63\% and 82\% of the measurements match each other within an interval of $1\rm \sigma$ and $2\rm \sigma$, respectively, pointing out to an underestimation of the uncertainties in the $F^{\rm J-PLUS}_{\rm H\alpha}$ values. In order to look deeper into this subject, we compute the variable $\Delta_{\rm F}$, defined as
\begin{equation}
 \Delta_{\rm F} = \frac{F^{\rm J-PLUS}_{\rm H\alpha}-F^{\rm spec}_{\rm H\alpha}}{\rm \sigma_{F^{\rm J-PLUS}_{\rm H\alpha}}} .
\end{equation}
If the $\rm \sigma_{F^{\rm J-PLUS}_{\rm H\alpha}}$ values are good descriptors of the accuracy of the measurements, the values of $\Delta_{\rm F}$ should lie in a normal distribution centered around zero with a unit dispersion \citep[e.g.][]{ilbert,carrasco,clsj14}. When this distribution is computed, the $\Delta_{\rm F}$ values of the measurements lie normally centered around zero, but the dispersion is $1.29$, confirming the underestimation of the uncertainties.

We also observe that the dispersion found in the present work is larger than the one in \citet{Gonzalo15}. When looking at the individual comparison of the photometric data with each spectroscopic dataset in Fig.~\ref{figure6}, offsets in opposite directions are noticeable. One could think that the increase in the dispersion could be caused by the combination of these individual comparisons. However, the median comparison ratio of each spectroscopic dataset is $\mathcal{R} = 1.10 \pm 0.25$ for SDSS and $\mathcal{R} = 0.90 \pm 0.36$ for CALIFA; being the dispersion similar in both cases. We have to take into account that in the present work, not only spectroscopic vs photometric observations are being compared, but also all the procedures and codes applied from observations to the final ${\rm H}\alpha$ emission line flux result, such as reduction and calibration processes, continuum fitting with differents SSPs or emission line modeling among others. All these steps involved, can cause the observed increase in the dispersion with respect to the value given by \citet{Gonzalo15}, done with synthetic J-PLUS photometry.

For this same reason, the systematic uncertainty of 15\% in the measurements derived by \citet{Gonzalo15} remains insufficient and needs to be updated to account for the observed uncertainties when the comparison is performed with real data. In fact, by setting a systematic uncertainty of 20\%, the $\Delta_{\rm F}$ distribution has a unity dispersion, and 67\% and 90\% of the measurements match each other within an interval of $1\rm \sigma$ and $2\rm \sigma$ respectively. This new derived error budget is representative of the uncertainties in the measurements.

The results show that the methodology presented in \citet{Gonzalo15} allows the exploitation of the J-PLUS photometric data in star formation studies in the nearby universe, by providing reliable ${\rm H}\alpha$ emission line fluxes and uncertainties.

\section{Spatially resolved SFR of NGC3995 and NGC3994}\label{science}

It has been reported that galaxies in close interacting pairs can show an enhancement of the SFR in their central parts up to a factor of ten \citep{patton,scuder}, being higher as the separation of the pair decreases. This enhancement is interpreted as an effect of the turbulence of the infalling gas. The previous studies could not measure ${\rm H}\alpha$ emission beyond the centers of the galaxies, since they were carried out with spectroscopic single fibers. Recently, \citet{cortijo} observed an enhancement of the SFR in the central parts and the disk of early-stage mergers with CALIFA data. In the J-PLUS SVD 1500041 fields observed for this work, there is a sample of nearby star-forming galaxies. This enables us to study the 2D star formation properties of two of them as an example of the J-PLUS potential. More specifically, we selected a close pair of interacting galaxies in the initial phase of the merging process. NGC3995 and NGC3994 in the triplet Arp313 at a projected distance of ${\rm r}_{\rm p} = 17.1h^{-1} \ {\rm kpc}$ \citep{Barton}. 

The objective is to get SFR radial profiles and 2D SFR maps of each galaxy. In order to do so, we started with the PSF homogenized images of the field, and then used \texttt{SExtractor} \citep{Bertin3} to retrieve morphological information of the galaxies. For that, the $r-$band image was used with a configuration that allows the code to recognize large galaxies as a single source. The \texttt{SExtractor} main configuration parameters to achieve this goal are the DEBLEND\_NTHRESH and both the DETECT\_MAXAREA and DETECT\_MINAREA, that we tuned to get large source detections. In this way, the position, ellipticity, angle of the semi-major axis with respect to the north and half light radius of the galaxies were measured. Next, aperture photometry on 30 elliptical apertures of increasing radius was performed  on the galaxies from the center to three half light radii extent. By making the aperture photometry in terms of the half light radius ($\rm R/R_{\rm eff}$), a common physical scale that allows the comparison of galaxies of different sizes is established. The flux and the error for every band in each aperture were calculated as previously explained in Sect.~\ref{eqs}. The SED-fitting method and the empirical relations were applied to the resulting SED to get the ${\rm H}\alpha$ flux for every elliptical aperture. Redshift based \citep{redshifts} luminosity distances were computed, obtaining ${\rm H}\alpha$ luminosities. The SFR of every aperture was estimated with the calibration given by \citet{kenicut} for a \citet{salpeter} initial mass function (IMF),
\begin{equation}
{\rm SFR} \ = \ 7.09\times10^{-42}L_{\rm H\alpha}\ \ \ \ {\rm [}M_{\odot}\,{\rm yr}^{-1}{\rm ]}.
\end{equation}

\begin{figure*}
\resizebox{\hsize}{!}{\includegraphics{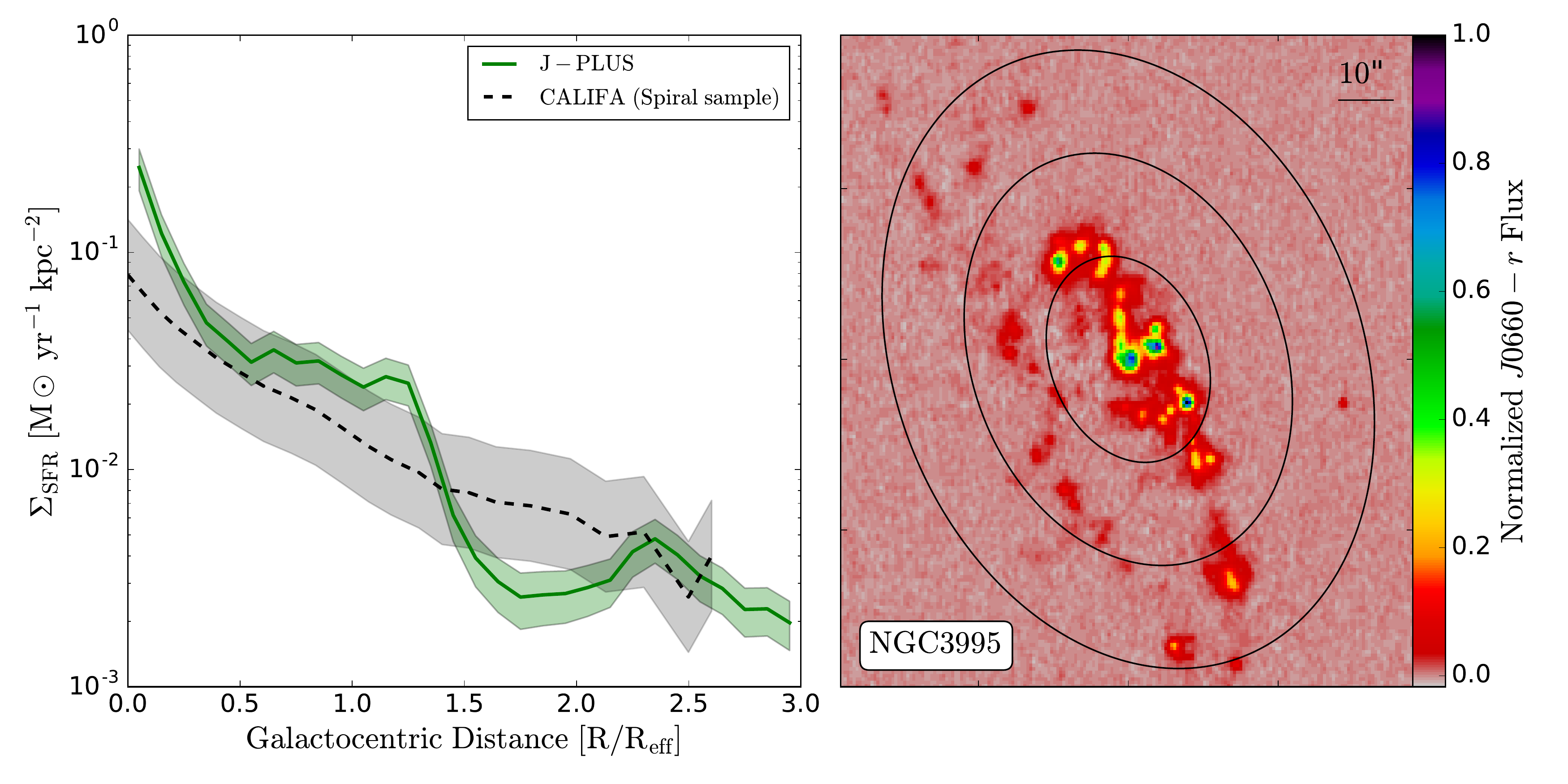}}
\resizebox{\hsize}{!}{\includegraphics{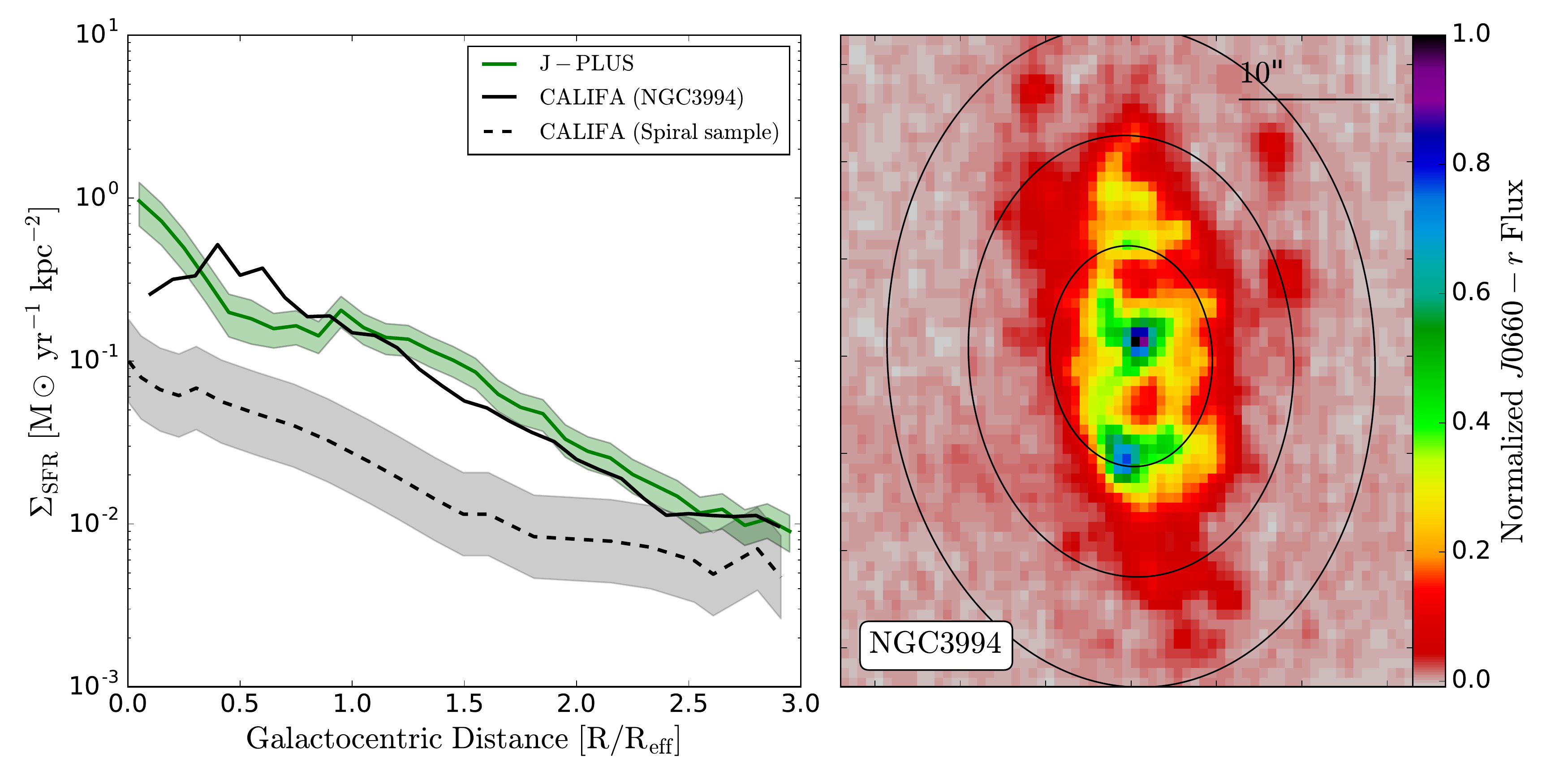}}
\caption{Star formation rate surface density (${\rm \Sigma_{SFR}}$) with the shaded area representing the $1\rm \sigma$ uncertainty of NGC3995 ({\it top left panel}) and NGC3994 ({\it bottom left panel}). 2D maps of the excess in the $J0660$ filter ({\it right panels}), representing the ${\rm H}\alpha$ emission areas are shown. One, two and three $\rm R/R_{\rm eff}$ ellipses are overplotted. ${\rm \Sigma_{SFR}}$ are compared in each case with a control sample of spirals matched in stellar mass and morphology (dashed black lines). Grey shaded areas show the dispersion of ${\rm \Sigma_{SFR}}$ in the spiral samples. The solid black line ({\it bottom left panel}) shows the CALIFA ${\rm \Sigma_{SFR}}$ measurement of the galaxy. Both the spiral samples and the individual galaxy measurements were performed by \citet{Rosa}.}
\label{figure9}
\end{figure*}

Finally, the SFR surface density (${\rm \Sigma_{SFR}}$) was estimated dividing the SFR by the area of each elliptical aperture. The results are commented in the next subsections and presented in Fig.~\ref{figure9}.

\subsection{Total star formation rate}
NGC3995 is too big to fit in the FoV of IFUs of surveys such as CALIFA or MaNGA, but not for J-PLUS. We measured a star formation rate of ${\rm SFR}=4.2 \pm 0.9$ for this galaxy. he estimate by the MPA-JHU group \citep{MPA}, is ${\rm SFR}=4.8 \pm 4.2$, made with a combination of spectroscopic and photometric data from SDSS that induces large uncertainties.

NGC3994 is a Sbc spiral \citep{CALIFA} with smaller size, and is part of the sample analyzed by CALIFA.	We calculated its total star formation rate to be ${\rm SFR}=2.9 \pm 0.6$. There are several estimates of the total SFR of this galaxy based on the same CALIFA data: \citet{catalan} give ${\rm SFR}=5.6 \pm 1.4$, \citet{Sanchez2} measure ${\rm SFR}=3.7 \pm 0.4$, \citet{cano} report ${\rm SFR}=4.5 \pm 1.5$, and \citet{catalan2} recently estimate ${\rm SFR}=3.16 \pm 0.12$. The estimate given by the MPA-JHU group is ${\rm SFR}=1.6 \pm 1.3$.

J-PLUS ${\rm SFR}$ estimates for NGC3994 and NGC3995 are compatible with the ones in the literature within 1$\sigma$, supporting both the developed methodologies and the quality of the J-PLUS dataset. All the ${\rm SFR}$ values given have been scaled to a common \citet{salpeter} IMF.			

\subsection{Star formation rate surface density}
With all the gathered information, we proceeded to study the ${\rm \Sigma_{SFR}}$ of NGC3995 and NGC3994 (Fig.~\ref{figure9}).

In the case of NGC3995 ({\it upper panel} in Fig.~\ref{figure9}), the ${\rm \Sigma_{SFR}}$ decreases as the galactocentric distance increases. In order to estimate the impact of the interaction in the star formation of NGC3995, we compared our ${\rm \Sigma_{SFR}}$ measurements with a control sample of non-interacting galaxies matched in morphology and stellar mass from \citet{Rosa}. The stellar mass of NGC3995 was calculated using the empirical relationship given by \citet{masas} based on the ($g$ - $i$) color of the galaxy, giving a value of $\rm{log \ M_{\star} \ [M_{\odot}]} = 9.52$. We found that a ${\rm \Sigma_{SFR}}$ enhancement is present not only in the central part of the galaxy but also around 1 $\rm R/R_{\rm eff}$, in addition, a decrement is observed between 1.4 and 2.4 $\rm R/R_{\rm eff}$. The enhancement is not conclusive, since it is compatible within 2$\sigma$ with the dispersion in the ${\rm \Sigma_{SFR}}$ control sample by \citet{Rosa}. The abrupt fall observed is most likely related to the distorsion induced by NGC3994 on the galaxy, which is more evident at the same $\rm R/R_{\rm eff}$ in the ${\rm H}\alpha$ excess image, with an assymetry in the distribution of the knots of star formation.

In the case of NGC3994 ({\it bottom panel} in Fig.~\ref{figure9}), we also observe how ${\rm \Sigma_{SFR}}$ decreases as the galactocentric distance increases. The derived values for this galaxy are consistent with those found by \citet{Rosa} using CALIFA data with a different methodology, that consists in estimating the fraction of stellar population younger than $\sim 30 \ {\rm Myr}$ from the spatially-resolved Star Formation History (SFH) of the galaxy, derived by the full spectral synthesis of the data cube (see \citealt{Rosa} for details). As in the previous case, we compared our ${\rm \Sigma_{SFR}}$ with a a control sample of non-interacting galaxies matched in morphology and stellar mass from \citet{Rosa}. The stellar mass of NGC3994 was calculated, giving a value of $\rm{log \ M_{\star} \ [M_{\odot}]} = 10.41$. In this case, we found a ${\rm \Sigma_{SFR}}$ enhancement in the entire disk of the galaxy and not only in the central part.

This is an illustrative example of the science that can be adressed with J-PLUS, carrying out simultaneously spatially resolved and environmental studies in galaxies of all morphological types and sizes in the nearby universe.

\section{Summary and conclusions}\label{summary}

We have tested the capabilities of measuring the ${\rm H}\alpha$ emission line flux from J-PLUS photometric data by comparing J-PLUS and spectroscopic measurements from SDSS and CALIFA in 46 star-forming regions. We have described the selection and how to extract the ${\rm H}\alpha$ emission line fluxes from them in the different datasets. We applied the SED-fitting technique and the empirical relations developed in \citet{Gonzalo15} to extract the $\rm [NII]$ free and dust corrected ${\rm H}\alpha$ fluxes, in the case of J-PLUS photometric data. We have performed a comparison between the photometric and spectroscopic ${\rm H}\alpha$ emission line fluxes, yielding a median ratio of $\mathcal{R}=1.05 \pm 0.25$ and confirming the expectations from \citet{Gonzalo15} with J-PLUS simulated data.

This result validates the developed methodology to extract the ${\rm H}\alpha$ flux from J-PLUS data. It provides an unbiased estimator of ${\rm H}\alpha$ emission line flux, with an error budget representative of the statistical and systematic uncertainties in the measurements. The degree of agreement between the compared fluxes is remarkable, given that we are dealing with photometric and spectroscopic measurements along with their different codes and procedures. The developed methodologies can be also applied to upcoming narrow band photometric surveys such as S-PLUS and J-PAS. 

Finally, we have studied the properties of the 2D star formation in a close pair of interacting galaxies, NGC3994 and NGC3995; finding an enhancement of the SFR in the center and outer parts of the disk of NGC3994, and obtaining total SFR estimates compatible with values in the literature. It is important to mention that J-PLUS, as a photometric survey, suffers from a lack of spectral resolution. In particular, the restricted number of narrow filters limits the number of emission lines that can be measured. This is important to detect zones in which AGN activity is the main ionizacion source via diagnosis diagrams, and can affect the ${\rm H}\alpha$ flux estimates in those areas. These and other issues imposed by the spectral resolution will be further explored in future publications. 

Once J-PLUS has completed its planned footprint, we expect to have data for approximately 5,000 galaxies at $\rm z \le 0.015$; a highly suitable dataset to study several science cases related to the star formation in the nearby universe. This represents a significant increase with respect to IFS surveys as  MaNGA, CALIFA, SAMI or VENGA with 524, 251, 127 and 30 galaxies at the same redshift range respectively, 

\begin{acknowledgements}
This work is based on observations made with the JAST/T80 telescope at the Observatorio Astrof\'{\i}sico de
Javalambre (OAJ), in Teruel, owned, managed and operated by the Centro de Estudios de F\'{\i}sica del
Cosmos de Arag\'on. We acknowledge the OAJ Data Processing and Archiving Unit
(UPAD) for reducing and calibrating the OAJ data used in this work.
Funding for the J-PLUS Project has been provided by
the Governments of Spain and Arag\'on through the Fondo de Inversiones
de Teruel, the Arag\'on Government through the Reseach Groups E96 and E103,
the Spanish Ministry of Economy and Competitiveness (MINECO; under
grants AYA2015-66211-C2-1-P, AYA2015-66211-C2-2, AYA2012-30789 and ICTS-2009-14),
and European FEDER funding (FCDD10-4E-867, FCDD13-4E-2685). 
R.L.G acknowledges support from ``Obra social de la fundación bancaria Ibercaja ''. 
K.V. acknowledges the {\it Juan de la Cierva Incorporaci\'on} fellowship, IJCI-2014-21960, of the Spanish government.
R.A.D acknowledges support from CNPq through BP grant  312307/2015-2, CSIC through grant COOPB20263, FINEP grants  REF. 1217/13 - 01.13.0279.00 and REF 0859/10 - 01.10.0663.00 for partial hardware support for the J-PLUS project through the National Observatory of Brazil.
L.G. was supported in part by the US National Science Foundation under Grant AST-1311862.
R.M.G.D was supported by AYA2016-77846-P, AYA2014-57490-P, and Junta de Andalucía FQ1580.  
J.A.H.J. and S. A. thank to Brazilian institution CNPq for financial support through postdoctoral fellowship (project 150237/2017-0 and 300336/2016-0, respectively).
R.L.O. was partially supported by the Brazilian agency CNPq (Universal Grants 459553/2014-3, PQ 302037/2015-2, and PDE 200289/2017-9).
R.L.G wants to thank his thesis directors and CEFCA staff for all the teaching, support and feedback received during the development of this work; to the J-PLUS collaboration for the feedback received; to the SELGIFS collaboration (http://astro.ft.uam.es/selgifs/) for the training in IFS data management, and to R.M. González Delgado for the data provided for the comparison in Fig. ~\ref{figure9}.
This research made use of \texttt{AstroPy} \citep{astropy}, \texttt{Matplotlib} \citep{matplotlib} and \texttt{NumPy}, part of the \texttt{Python} Software Foundation. (Python Language Reference, version 2.7. Available at http://www.python.org).
This study uses data provided by the Calar Alto Legacy Integral Field Area (CALIFA) survey (http://califa.caha.es/). Based on observations collected at the Centro Astronómico Hispano Alemán (CAHA) at Calar Alto, operated jointly by the Max-Planck-Institut fűr Astronomie and the Instituto de Astrofísica de Andalucía (CSIC). 
This study uses data provided by SDSS. Funding for SDSS-III has been provided by the Alfred P. Sloan Foundation, the Participating Institutions, the National Science Foundation, and the U.S. Department of Energy Office of Science. The SDSS-III web site is http://www.sdss3.org/. SDSS-III is managed by the Astrophysical Research Consortium for the Participating Institutions of the SDSS-III Collaboration including the University of Arizona, the Brazilian Participation Group, Brookhaven National Laboratory, Carnegie Mellon University, University of Florida, the French Participation Group, the German Participation Group, Harvard University, the Instituto de Astrofisica de Canarias, the Michigan State/Notre Dame/JINA Participation Group, Johns Hopkins University, Lawrence Berkeley National Laboratory, Max Planck Institute for Astrophysics, Max Planck Institute for Extraterrestrial Physics, New Mexico State University, New York University, Ohio State University, Pennsylvania State University, University of Portsmouth, Princeton University, the Spanish Participation Group, University of Tokyo, University of Utah, Vanderbilt University, University of Virginia, University of Washington, and Yale University.
This research has made use of the NASA/IPAC Extragalactic Database (NED) which is operated by the Jet Propulsion Laboratory, California Institute of Technology, under contract with the National Aeronautics and Space Administration. This research has made use of the SIMBAD database, operated at CDS, Strasbourg, France \citep{simbad}.
\end{acknowledgements}

\bibliographystyle{aa} 
\bibliography{mybib}

\begin{thebibliography}{74}
\expandafter\ifx\csname natexlab\endcsname\relax\def\natexlab#1{#1}\fi

\bibitem[{{Alam} {et~al.}(2015){Alam}, {Albareti}, {Allende Prieto}, {Anders},
  {Anderson}, {Anderton}, {Andrews}, {Armengaud}, {Aubourg}, {Bailey}, \&
  et~al.}]{dr12}
{Alam}, S., {Albareti}, F.~D., {Allende Prieto}, C., {et~al.} 2015, \apjs, 219,
  12

\bibitem[{{Astropy Collaboration} {et~al.}(2013){Astropy Collaboration},
  {Robitaille}, {Tollerud}, {Greenfield}, {Droettboom}, {Bray}, {Aldcroft},
  {Davis}, {Ginsburg}, {Price-Whelan}, {Kerzendorf}, {Conley}, {Crighton},
  {Barbary}, {Muna}, {Ferguson}, {Grollier}, {Parikh}, {Nair}, {Unther},
  {Deil}, {Woillez}, {Conseil}, {Kramer}, {Turner}, {Singer}, {Fox}, {Weaver},
  {Zabalza}, {Edwards}, {Azalee Bostroem}, {Burke}, {Casey}, {Crawford},
  {Dencheva}, {Ely}, {Jenness}, {Labrie}, {Lim}, {Pierfederici}, {Pontzen},
  {Ptak}, {Refsdal}, {Servillat}, \& {Streicher}}]{astropy}
{Astropy Collaboration}, {Robitaille}, T.~P., {Tollerud}, E.~J., {et~al.} 2013,
  \aap, 558, A33

\bibitem[{{Bacon} {et~al.}(2010){Bacon}, {Accardo}, {Adjali}, {Anwand},
  {Bauer}, {Biswas}, {Blaizot}, {Boudon}, {Brau-Nogue}, {Brinchmann},
  {Caillier}, {Capoani}, {Carollo}, {Contini}, {Couderc}, {Daguis{\'e}},
  {Deiries}, {Delabre}, {Dreizler}, {Dubois}, {Dupieux}, {Dupuy}, {Emsellem},
  {Fechner}, {Fleischmann}, {Fran{\c c}ois}, {Gallou}, {Gharsa}, {Glindemann},
  {Gojak}, {Guiderdoni}, {Hansali}, {Hahn}, {Jarno}, {Kelz}, {Koehler},
  {Kosmalski}, {Laurent}, {Le Floch}, {Lilly}, {Lizon}, {Loupias}, {Manescau},
  {Monstein}, {Nicklas}, {Olaya}, {Pares}, {Pasquini}, {P{\'e}contal-Rousset},
  {Pell{\'o}}, {Petit}, {Popow}, {Reiss}, {Remillieux}, {Renault}, {Roth},
  {Rupprecht}, {Serre}, {Schaye}, {Soucail}, {Steinmetz}, {Streicher}, {Stuik},
  {Valentin}, {Vernet}, {Weilbacher}, {Wisotzki}, \& {Yerle}}]{MUSE}
{Bacon}, R., {Accardo}, M., {Adjali}, L., {et~al.} 2010, in \procspie, Vol.
  7735, Ground-based and Airborne Instrumentation for Astronomy III, 773508

\bibitem[{{Bacon} {et~al.}(2001){Bacon}, {Copin}, {Monnet}, {Miller},
  {Allington-Smith}, {Bureau}, {Carollo}, {Davies}, {Emsellem}, {Kuntschner},
  {Peletier}, {Verolme}, \& {de Zeeuw}}]{SAURON}
{Bacon}, R., {Copin}, Y., {Monnet}, G., {et~al.} 2001, \mnras, 326, 23

\bibitem[{{Baldwin} {et~al.}(1981){Baldwin}, {Phillips}, \& {Terlevich}}]{BPT}
{Baldwin}, J.~A., {Phillips}, M.~M., \& {Terlevich}, R. 1981, \pasp, 93, 5

\bibitem[{{Barrera-Ballesteros} {et~al.}(2015){Barrera-Ballesteros},
  {S{\'a}nchez}, {Garc{\'{\i}}a-Lorenzo}, {Falc{\'o}n-Barroso}, {Mast},
  {Garc{\'{\i}}a-Benito}, {Husemann}, {van de Ven}, {Iglesias-P{\'a}ramo},
  {Rosales-Ortega}, {P{\'e}rez-Torres}, {M{\'a}rquez}, {Kehrig}, {Marino},
  {Vilchez}, {Galbany}, {L{\'o}pez-S{\'a}nchez}, {Walcher}, \& {Califa
  Collaboration}}]{barrera}
{Barrera-Ballesteros}, J.~K., {S{\'a}nchez}, S.~F., {Garc{\'{\i}}a-Lorenzo},
  B., {et~al.} 2015, \aap, 579, A45

\bibitem[{{Barton Gillespie} {et~al.}(2003){Barton Gillespie}, {Geller}, \&
  {Kenyon}}]{Barton}
{Barton Gillespie}, E., {Geller}, M.~J., \& {Kenyon}, S.~J. 2003, \apj, 582,
  668

\bibitem[{{Benitez} {et~al.}(2014){Benitez}, {Dupke}, {Moles}, {Sodre},
  {Cenarro}, {Marin-Franch}, {Taylor}, {Cristobal}, {Fernandez-Soto}, {Mendes
  de Oliveira}, {Cepa-Nogue}, {Abramo}, {Alcaniz}, {Overzier},
  {Hernandez-Monteagudo}, {Alfaro}, {Kanaan}, {Carvano}, {Reis}, {Martinez
  Gonzalez}, {Ascaso}, {Ballesteros}, {Xavier}, {Varela}, {Ederoclite},
  {Vazquez Ramio}, {Broadhurst}, {Cypriano}, {Angulo}, {Diego}, {Zandivarez},
  {Diaz}, {Melchior}, {Umetsu}, {Spinelli}, {Zitrin}, {Coe}, {Yepes}, {Vielva},
  {Sahni}, {Marcos-Caballero}, {Shu Kitaura}, {Maroto}, {Masip}, {Tsujikawa},
  {Carneiro}, {Gonzalez Nuevo}, {Carvalho}, {Reboucas}, {Carvalho}, {Abdalla},
  {Bernui}, {Pigozzo}, {Ferreira}, {Chandrachani Devi}, {Bengaly}, {Campista},
  {Amorim}, {Asari}, {Bongiovanni}, {Bonoli}, {Bruzual}, {Cardiel}, {Cava},
  {Cid Fernandes}, {Coelho}, {Cortesi}, {Delgado}, {Diaz Garcia}, {Espinosa},
  {Galliano}, {Gonzalez-Serrano}, {Falcon-Barroso}, {Fritz}, {Fernandes},
  {Gorgas}, {Hoyos}, {Jimenez-Teja}, {Lopez-Aguerri}, {Lopez-San Juan},
  {Mateus}, {Molino}, {Novais}, {OMill}, {Oteo}, {Perez-Gonzalez}, {Poggianti},
  {Proctor}, {Ricciardelli}, {Sanchez-Blazquez}, {Storchi-Bergmann}, {Telles},
  {Schoennell}, {Trujillo}, {Vazdekis}, {Viironen}, {Daflon},
  {Aparicio-Villegas}, {Rocha}, {Ribeiro}, {Borges}, {Martins}, {Marcolino},
  {Martinez-Delgado}, {Perez-Torres}, {Siffert}, {Calvao}, {Sako}, {Kessler},
  {Alvarez-Candal}, {De Pra}, {Roig}, {Lazzaro}, {Gorosabel}, {Lopes de
  Oliveira}, {Lima-Neto}, {Irwin}, {Liu}, {Alvarez}, {Balmes}, {Chueca},
  {Costa-Duarte}, {da Costa}, {Dantas}, {Diaz}, {Fabregat}, {Ferrari},
  {Gavela}, {Gracia}, {Gruel}, {Gutierrez}, {Guzman}, {Hernandez-Fernandez},
  {Herranz}, {Hurtado-Gil}, {Jablonsky}, {Laporte}, {Le Tiran}, {Licandro},
  {Lima}, {Martin}, {Martinez}, {Montero}, {Penteado}, {Pereira}, {Peris},
  {Quilis}, {Sanchez-Portal}, {Soja}, {Solano}, {Torra}, \&
  {Valdivielso}}]{Benitez}
{Benitez}, N., {Dupke}, R., {Moles}, M., {et~al.} 2014, ArXiv e-prints
  [\eprint[arXiv]{1403.5237}]

\bibitem[{{Bertin}(2011)}]{Bertin1}
{Bertin}, E. 2011, in Astronomical Society of the Pacific Conference Series,
  Vol. 442, Astronomical Data Analysis Software and Systems XX, ed. I.~N.
  {Evans}, A.~{Accomazzi}, D.~J. {Mink}, \& A.~H. {Rots}, 435

\bibitem[{{Bertin}(2013)}]{Bertin2}
{Bertin}, E. 2013, {PSFEx: Point Spread Function Extractor}, Astrophysics
  Source Code Library

\bibitem[{{Bertin} \& {Arnouts}(1996)}]{Bertin3}
{Bertin}, E. \& {Arnouts}, S. 1996, \aaps, 117, 393

\bibitem[{{Blanc} {et~al.}(2013){Blanc}, {Weinzirl}, {Song}, {Heiderman},
  {Gebhardt}, {Jogee}, {Evans}, {van den Bosch}, {Luo}, {Drory}, {Fabricius},
  {Fisher}, {Hao}, {Kaplan}, {Marinova}, {Vutisalchavakul}, \&
  {Yoachim}}]{VIRUS}
{Blanc}, G.~A., {Weinzirl}, T., {Song}, M., {et~al.} 2013, \aj, 145, 138

\bibitem[{{Bonatto} {et~al.}(2018){Bonatto}, {Chies-Santos}, {Coelho}, \&
  {J-PLUS collaboration}}]{bonatto18}
{Bonatto}, C., {Chies-Santos}, A.~L., {Coelho}, P.~R.~T., \& {J-PLUS
  collaboration}. 2018, \aap, in press

\bibitem[{{Bothwell} {et~al.}(2014){Bothwell}, {Kenicutt}, {Johnson}, {Wu},
  {Lee}, {Dale}, {Engelbracht}, {Calzetti}, \& {Skillman}}]{bothwell}
{Bothwell}, M.~S., {Kenicutt}, R.~C., {Johnson}, B.~D., {et~al.} 2014, \mnras,
  438, 3608

\bibitem[{{Brinchmann} {et~al.}(2004){Brinchmann}, {Charlot}, {White},
  {Tremonti}, {Kauffmann}, {Heckman}, \& {Brinkmann}}]{MPA}
{Brinchmann}, J., {Charlot}, S., {White}, S.~D.~M., {et~al.} 2004, \mnras, 351,
  1151

\bibitem[{{Bruzual} \& {Charlot}(2003)}]{Bruzual}
{Bruzual}, G. \& {Charlot}, S. 2003, \mnras, 344, 1000

\bibitem[{{Bundy} {et~al.}(2015){Bundy}, {Bershady}, {Law}, {Yan}, {Drory},
  {MacDonald}, {Wake}, {Cherinka}, {S{\'a}nchez-Gallego}, {Weijmans}, {Thomas},
  {Tremonti}, {Masters}, {Coccato}, {Diamond-Stanic}, {Arag{\'o}n-Salamanca},
  {Avila-Reese}, {Badenes}, {Falc{\'o}n-Barroso}, {Belfiore}, {Bizyaev},
  {Blanc}, {Bland-Hawthorn}, {Blanton}, {Brownstein}, {Byler}, {Cappellari},
  {Conroy}, {Dutton}, {Emsellem}, {Etherington}, {Frinchaboy}, {Fu}, {Gunn},
  {Harding}, {Johnston}, {Kauffmann}, {Kinemuchi}, {Klaene}, {Knapen},
  {Leauthaud}, {Li}, {Lin}, {Maiolino}, {Malanushenko}, {Malanushenko}, {Mao},
  {Maraston}, {McDermid}, {Merrifield}, {Nichol}, {Oravetz}, {Pan}, {Parejko},
  {Sanchez}, {Schlegel}, {Simmons}, {Steele}, {Steinmetz}, {Thanjavur},
  {Thompson}, {Tinker}, {van den Bosch}, {Westfall}, {Wilkinson}, {Wright},
  {Xiao}, \& {Zhang}}]{Bundy}
{Bundy}, K., {Bershady}, M.~A., {Law}, D.~R., {et~al.} 2015, \apj, 798, 7

\bibitem[{{Calzetti}(2013)}]{calzetti2}
{Calzetti}, D. 2013, {Star Formation Rate Indicators}, ed.
  J.~{Falc{\'o}n-Barroso} \& J.~H. {Knapen}, 419

\bibitem[{{Calzetti} {et~al.}(2000){Calzetti}, {Armus}, {Bohlin}, {Kinney},
  {Koornneef}, \& {Storchi-Bergmann}}]{calzetti}
{Calzetti}, D., {Armus}, L., {Bohlin}, R.~C., {et~al.} 2000, \apj, 533, 682

\bibitem[{{Cano-D{\'{\i}}az} {et~al.}(2016){Cano-D{\'{\i}}az}, {S{\'a}nchez},
  {Zibetti}, {Ascasibar}, {Bland-Hawthorn}, {Ziegler}, {Gonz{\'a}lez Delgado},
  {Walcher}, {Garc{\'{\i}}a-Benito}, {Mast}, {Mendoza-P{\'e}rez},
  {Falc{\'o}n-Barroso}, {Galbany}, {Husemann}, {Kehrig}, {Marino},
  {S{\'a}nchez-Bl{\'a}zquez}, {L{\'o}pez-Cob{\'a}}, {L{\'o}pez-S{\'a}nchez}, \&
  {Vilchez}}]{cano}
{Cano-D{\'{\i}}az}, M., {S{\'a}nchez}, S.~F., {Zibetti}, S., {et~al.} 2016,
  \apjl, 821, L26

\bibitem[{{Cappellari} {et~al.}(2011){Cappellari}, {Emsellem}, {Krajnovi{\'c}},
  {McDermid}, {Scott}, {Verdoes Kleijn}, {Young}, {Alatalo}, {Bacon}, {Blitz},
  {Bois}, {Bournaud}, {Bureau}, {Davies}, {Davis}, {de Zeeuw}, {Duc},
  {Khochfar}, {Kuntschner}, {Lablanche}, {Morganti}, {Naab}, {Oosterloo},
  {Sarzi}, {Serra}, \& {Weijmans}}]{Cappellari}
{Cappellari}, M., {Emsellem}, E., {Krajnovi{\'c}}, D., {et~al.} 2011, \mnras,
  413, 813

\bibitem[{{Carrasco Kind} \& {Brunner}(2013)}]{carrasco}
{Carrasco Kind}, M. \& {Brunner}, R.~J. 2013, \mnras, 432, 1483

\bibitem[{{Catal{\'a}n-Torrecilla} {et~al.}(2015){Catal{\'a}n-Torrecilla}, {Gil
  de Paz}, {Castillo-Morales}, {Iglesias-P{\'a}ramo}, {S{\'a}nchez},
  {Kennicutt}, {P{\'e}rez-Gonz{\'a}lez}, {Marino}, {Walcher}, {Husemann},
  {Garc{\'{\i}}a-Benito}, {Mast}, {Gonz{\'a}lez Delgado}, {Mu{\~n}oz-Mateos},
  {Bland-Hawthorn}, {Bomans}, {Del Olmo}, {Galbany}, {Gomes}, {Kehrig},
  {L{\'o}pez-S{\'a}nchez}, {Mendoza}, {Monreal-Ibero}, {P{\'e}rez-Torres},
  {S{\'a}nchez-Bl{\'a}zquez}, {Vilchez}, \& {Califa Collaboration}}]{catalan}
{Catal{\'a}n-Torrecilla}, C., {Gil de Paz}, A., {Castillo-Morales}, A.,
  {et~al.} 2015, \aap, 584, A87

\bibitem[{{Catal{\'a}n-Torrecilla} {et~al.}(2017){Catal{\'a}n-Torrecilla}, {Gil
  de Paz}, {Castillo-Morales}, {M{\'e}ndez-Abreu}, {Falc{\'o}n-Barroso},
  {Bekeraite}, {Costantin}, {de Lorenzo-C{\'a}ceres}, {Florido},
  {Garc{\'{\i}}a-Benito}, {Husemann}, {Iglesias-P{\'a}ramo}, {Kennicutt},
  {Mast}, {Pascual}, {Ruiz-Lara}, {S{\'a}nchez-Menguiano}, {S{\'a}nchez},
  {Walcher}, {Bland-Hawthorn}, {Duarte Puertas}, {Marino}, {Masegosa},
  {S{\'a}nchez-Bl{\'a}zquez}, \& {CALIFA Collaboration}}]{catalan2}
{Catal{\'a}n-Torrecilla}, C., {Gil de Paz}, A., {Castillo-Morales}, A.,
  {et~al.} 2017, ArXiv e-prints [\eprint[arXiv]{1709.01035}]

\bibitem[{{Cenarro} {et~al.}(2018){Cenarro}, {Crist{\'o}bal-Hornillos},
  {Mar{\'{\i}}n-Franch}, \& {J-PLUS collaboration}}]{cenarro18}
{Cenarro}, A.~J., {Crist{\'o}bal-Hornillos}, D., {Mar{\'{\i}}n-Franch}, A., \&
  {J-PLUS collaboration}. 2018, \aap, submitted [arXiv:1804.02667]

\bibitem[{{Cortijo-Ferrero} {et~al.}(2017){Cortijo-Ferrero}, {Gonz{\'a}lez
  Delgado}, {P{\'e}rez}, {Cid Fernandes}, {Garc{\'{\i}}a-Benito}, {Di Matteo},
  {S{\'a}nchez}, {de Amorim}, {Lacerda}, {L{\'o}pez Fern{\'a}ndez}, \&
  {Tadhunter}}]{cortijo}
{Cortijo-Ferrero}, C., {Gonz{\'a}lez Delgado}, R.~M., {P{\'e}rez}, E., {et~al.}
  2017, \aap, 607, A70

\bibitem[{{Croom} {et~al.}(2012){Croom}, {Lawrence}, {Bland-Hawthorn},
  {Bryant}, {Fogarty}, {Richards}, {Goodwin}, {Farrell}, {Miziarski}, {Heald},
  {Jones}, {Lee}, {Colless}, {Brough}, {Hopkins}, {Bauer}, {Birchall}, {Ellis},
  {Horton}, {Leon-Saval}, {Lewis}, {L{\'o}pez-S{\'a}nchez}, {Min}, {Trinh}, \&
  {Trowland}}]{Croom}
{Croom}, S.~M., {Lawrence}, J.~S., {Bland-Hawthorn}, J., {et~al.} 2012, \mnras,
  421, 872

\bibitem[{{Drory} {et~al.}(2015){Drory}, {MacDonald}, {Bershady}, {Bundy},
  {Gunn}, {Law}, {Smith}, {Stoll}, {Tremonti}, {Wake}, {Yan}, {Weijmans},
  {Byler}, {Cherinka}, {Cope}, {Eigenbrot}, {Harding}, {Holder}, {Huehnerhoff},
  {Jaehnig}, {Jansen}, {Klaene}, {Paat}, {Percival}, \& {Sayres}}]{mangaifu}
{Drory}, N., {MacDonald}, N., {Bershady}, M.~A., {et~al.} 2015, \aj, 149, 77

\bibitem[{{Falc{\'o}n-Barroso} {et~al.}(2011){Falc{\'o}n-Barroso},
  {S{\'a}nchez-Bl{\'a}zquez}, {Vazdekis}, {Ricciardelli}, {Cardiel}, {Cenarro},
  {Gorgas}, \& {Peletier}}]{falcon}
{Falc{\'o}n-Barroso}, J., {S{\'a}nchez-Bl{\'a}zquez}, P., {Vazdekis}, A.,
  {et~al.} 2011, \aap, 532, A95

\bibitem[{{Gallego} {et~al.}(1996){Gallego}, {Zamorano}, {Aragon-Salamanca}, \&
  {Rego}}]{gallego}
{Gallego}, J., {Zamorano}, J., {Aragon-Salamanca}, A., \& {Rego}, M. 1996,
  \apjl, 459, L43

\bibitem[{{Garc{\'{\i}}a-Benito} {et~al.}(2015){Garc{\'{\i}}a-Benito},
  {Zibetti}, {S{\'a}nchez}, {Husemann}, {de Amorim}, {Castillo-Morales}, {Cid
  Fernandes}, {Ellis}, {Falc{\'o}n-Barroso}, {Galbany}, {Gil de Paz},
  {Gonz{\'a}lez Delgado}, {Lacerda}, {L{\'o}pez-Fernandez}, {de
  Lorenzo-C{\'a}ceres}, {Lyubenova}, {Marino}, {Mast}, {Mendoza}, {P{\'e}rez},
  {Vale Asari}, {Aguerri}, {Ascasibar}, {Bekerait*error*{\.e}},
  {Bland-Hawthorn}, {Barrera-Ballesteros}, {Bomans}, {Cano-D{\'{\i}}az},
  {Catal{\'a}n-Torrecilla}, {Cortijo}, {Delgado-Inglada}, {Demleitner},
  {Dettmar}, {D{\'{\i}}az}, {Florido}, {Gallazzi}, {Garc{\'{\i}}a-Lorenzo},
  {Gomes}, {Holmes}, {Iglesias-P{\'a}ramo}, {Jahnke}, {Kalinova}, {Kehrig},
  {Kennicutt}, {L{\'o}pez-S{\'a}nchez}, {M{\'a}rquez}, {Masegosa}, {Meidt},
  {Mendez-Abreu}, {Moll{\'a}}, {Monreal-Ibero}, {Morisset}, {del Olmo},
  {Papaderos}, {P{\'e}rez}, {Quirrenbach}, {Rosales-Ortega}, {Roth},
  {Ruiz-Lara}, {S{\'a}nchez-Bl{\'a}zquez}, {S{\'a}nchez-Menguiano}, {Singh},
  {Spekkens}, {Stanishev}, {Torres-Papaqui}, {van de Ven}, {Vilchez},
  {Walcher}, {Wild}, {Wisotzki}, {Ziegler}, {Alves}, {Barrado}, {Quintana}, \&
  {Aceituno}}]{GarciaBenito}
{Garc{\'{\i}}a-Benito}, R., {Zibetti}, S., {S{\'a}nchez}, S.~F., {et~al.} 2015,
  \aap, 576, A135

\bibitem[{{Gil de Paz} {et~al.}(2016){Gil de Paz}, {Carrasco}, {Gallego},
  {Iglesias-P{\'a}ramo}, {Cedazo}, {Garc{\'{\i}}a Vargas}, {Arrillaga},
  {Avil{\'e}s}, {Cardiel}, {Carrera}, {Castillo-Morales},
  {Castillo-Dom{\'{\i}}nguez}, {de la Cruz Garc{\'{\i}}a}, {Esteban San
  Rom{\'a}n}, {Ferrusca}, {G{\'o}mez-{\'A}lvarez}, {Izazaga-P{\'e}rez},
  {Lefort}, {L{\'o}pez-Orozco}, {Maldonado}, {Mart{\'{\i}}nez-Delgado},
  {Morales Dur{\'a}n}, {Mujica}, {P{\'a}ez}, {Pascual}, {P{\'e}rez-Calpena},
  {Picazo}, {S{\'a}nchez-Penim}, {S{\'a}nchez-Blanco}, {Tulloch},
  {Vel{\'a}zquez}, {V{\'{\i}}lchez}, {Zamorano}, {Aguerri}, {Barrado y
  Nav{\'a}scues}, {Bertone}, {Cava}, {Cenarro}, {Ch{\'a}vez}, {Garc{\'{\i}}a},
  {Garc{\'{\i}}a-Rojas}, {Guichard}, {Gonz{\'a}lez-Delgado}, {Guzm{\'a}n},
  {Herrero}, {Hu{\'e}lamo}, {Hughes}, {Jim{\'e}nez-Vicente}, {Kehrig},
  {Marino}, {M{\'a}rquez}, {Masegosa}, {Mayya}, {M{\'e}ndez-Abreu},
  {Moll{\'a}}, {Mu{\~n}oz-Tu{\~n}{\'o}n}, {Peimbert}, {P{\'e}rez-Gonz{\'a}lez},
  {P{\'e}rez Montero}, {Rodr{\'{\i}}guez}, {Rodr{\'{\i}}guez-Espinosa},
  {Rodr{\'{\i}}guez-Merino}, {Rodr{\'{\i}}guez-Mu{\~n}oz}, {Rosa-Gonz{\'a}lez},
  {S{\'a}nchez-Almeida}, {S{\'a}nchez Contreras}, {S{\'a}nchez-Bl{\'a}zquez},
  {S{\'a}nchez Moreno}, {S{\'a}nchez}, {Sarajedini}, {Silich},
  {Sim{\'o}n-D{\'{\i}}az}, {Tenorio-Tagle}, {Terlevich}, {Terlevich},
  {Torres-Peimbert}, {Trujillo}, {Tsamis}, \& {Vega}}]{megara}
{Gil de Paz}, A., {Carrasco}, E., {Gallego}, J., {et~al.} 2016, in \procspie,
  Vol. 9908, Ground-based and Airborne Instrumentation for Astronomy VI, 99081K

\bibitem[{{Gonz{\'a}lez Delgado} {et~al.}(2016){Gonz{\'a}lez Delgado}, {Cid
  Fernandes}, {P{\'e}rez}, {Garc{\'{\i}}a-Benito}, {L{\'o}pez Fern{\'a}ndez},
  {Lacerda}, {Cortijo-Ferrero}, {de Amorim}, {Vale Asari}, {S{\'a}nchez},
  {Walcher}, {Wisotzki}, {Mast}, {Alves}, {Ascasibar}, {Bland-Hawthorn},
  {Galbany}, {Kennicutt}, {M{\'a}rquez}, {Masegosa}, {Moll{\'a}},
  {S{\'a}nchez-Bl{\'a}zquez}, \& {V{\'{\i}}lchez}}]{Rosa}
{Gonz{\'a}lez Delgado}, R.~M., {Cid Fernandes}, R., {P{\'e}rez}, E., {et~al.}
  2016, \aap, 590, A44

\bibitem[{Hunter(2007)}]{matplotlib}
Hunter, J.~D. 2007, Computing In Science \& Engineering, 9, 90

\bibitem[{{Ilbert} {et~al.}(2009){Ilbert}, {Capak}, {Salvato}, {Aussel},
  {McCracken}, {Sanders}, {Scoville}, {Kartaltepe}, {Arnouts}, {Le Floc'h},
  {Mobasher}, {Taniguchi}, {Lamareille}, {Leauthaud}, {Sasaki}, {Thompson},
  {Zamojski}, {Zamorani}, {Bardelli}, {Bolzonella}, {Bongiorno}, {Brusa},
  {Caputi}, {Carollo}, {Contini}, {Cook}, {Coppa}, {Cucciati}, {de la Torre},
  {de Ravel}, {Franzetti}, {Garilli}, {Hasinger}, {Iovino}, {Kampczyk},
  {Kneib}, {Knobel}, {Kovac}, {Le Borgne}, {Le Brun}, {Le F{\`e}vre}, {Lilly},
  {Looper}, {Maier}, {Mainieri}, {Mellier}, {Mignoli}, {Murayama}, {Pell{\`o}},
  {Peng}, {P{\'e}rez-Montero}, {Renzini}, {Ricciardelli}, {Schiminovich},
  {Scodeggio}, {Shioya}, {Silverman}, {Surace}, {Tanaka}, {Tasca}, {Tresse},
  {Vergani}, \& {Zucca}}]{ilbert}
{Ilbert}, O., {Capak}, P., {Salvato}, M., {et~al.} 2009, \apj, 690, 1236

\bibitem[{{Kauffmann} {et~al.}(2003){Kauffmann}, {Heckman}, {Tremonti},
  {Brinchmann}, {Charlot}, {White}, {Ridgway}, {Brinkmann}, {Fukugita}, {Hall},
  {Ivezi{\'c}}, {Richards}, \& {Schneider}}]{kauffmann}
{Kauffmann}, G., {Heckman}, T.~M., {Tremonti}, C., {et~al.} 2003, \mnras, 346,
  1055

\bibitem[{{Kennicutt} \& {Evans}(2012)}]{kenSFR}
{Kennicutt}, R.~C. \& {Evans}, N.~J. 2012, \araa, 50, 531

\bibitem[{{Kennicutt}(1998)}]{kenicut}
{Kennicutt}, Jr., R.~C. 1998, \araa, 36, 189

\bibitem[{{Kewley} {et~al.}(2001){Kewley}, {Dopita}, {Sutherland}, {Heisler},
  \& {Trevena}}]{kewley}
{Kewley}, L.~J., {Dopita}, M.~A., {Sutherland}, R.~S., {Heisler}, C.~A., \&
  {Trevena}, J. 2001, \apj, 556, 121

\bibitem[{{Labb{\'e}} {et~al.}(2003){Labb{\'e}}, {Franx}, {Rudnick},
  {Schreiber}, {Rix}, {Moorwood}, {van Dokkum}, {van der Werf},
  {R{\"o}ttgering}, {van Starkenburg}, {van der Wel}, {Kuijken}, \&
  {Daddi}}]{labbe}
{Labb{\'e}}, I., {Franx}, M., {Rudnick}, G., {et~al.} 2003, \aj, 125, 1107

\bibitem[{{L{\'o}pez-Sanjuan} {et~al.}(2014){L{\'o}pez-Sanjuan}, {Cenarro},
  {Hern{\'a}ndez-Monteagudo}, {Varela}, {Molino}, {Arnalte-Mur}, {Ascaso},
  {Castander}, {Fern{\'a}ndez-Soto}, {Huertas-Company}, {M{\'a}rquez},
  {Mart{\'{\i}}nez}, {Masegosa}, {Moles}, {Povi{\'c}}, {Aguerri}, {Alfaro},
  {Aparicio-Villegas}, {Ben{\'{\i}}tez}, {Broadhurst}, {Cabrera-Ca{\~n}o},
  {Cepa}, {Cervi{\~n}o}, {Crist{\'o}bal-Hornillos}, {Del Olmo}, {Gonz{\'a}lez
  Delgado}, {Husillos}, {Infante}, {Perea}, {Prada}, \& {Quintana}}]{clsj14}
{L{\'o}pez-Sanjuan}, C., {Cenarro}, A.~J., {Hern{\'a}ndez-Monteagudo}, C.,
  {et~al.} 2014, \aap, 564, A127

\bibitem[{{L{\'o}pez-Sanjuan} {et~al.}(2018){L{\'o}pez-Sanjuan},
  {V{\'a}zquez-Rami{\'o}}, {Varela}, \& {J-PLUS collaboration}}]{clsj18}
{L{\'o}pez-Sanjuan}, C., {V{\'a}zquez-Rami{\'o}}, H., {Varela}, J., \& {J-PLUS
  collaboration}. 2018, \aap, submitted [arXiv:1804.02673]

\bibitem[{{Mandel} {et~al.}(2001){Mandel}, {Murray}, \& {Roll}}]{funtools}
{Mandel}, E., {Murray}, S.~S., \& {Roll}, J.~B. 2001, in Astronomical Society
  of the Pacific Conference Series, Vol. 238, Astronomical Data Analysis
  Software and Systems X, ed. F.~R. {Harnden}, Jr., F.~A. {Primini}, \& H.~E.
  {Payne}, 225

\bibitem[{{Maraston} \& {Str{\"o}mb{\"a}ck}(2011)}]{maraston}
{Maraston}, C. \& {Str{\"o}mb{\"a}ck}, G. 2011, \mnras, 418, 2785

\bibitem[{{Marin-Franch} {et~al.}(2015){Marin-Franch}, {Taylor}, {Cenarro},
  {Cristobal-Hornillos}, \& {Moles}}]{Toni}
{Marin-Franch}, A., {Taylor}, K., {Cenarro}, J., {Cristobal-Hornillos}, D., \&
  {Moles}, M. 2015, IAU General Assembly, 22, 2257381

\bibitem[{{Mendes de Oliveira} \& {S-PLUS collaboration}(in prep.)}]{Claudia}
{Mendes de Oliveira}, C. \& {S-PLUS collaboration}. in prep.

\bibitem[{{Molino} {et~al.}(2014){Molino}, {Ben{\'{\i}}tez}, {Moles},
  {Fern{\'a}ndez-Soto}, {Crist{\'o}bal-Hornillos}, {Ascaso},
  {Jim{\'e}nez-Teja}, {Schoenell}, {Arnalte-Mur}, {Povi{\'c}}, {Coe},
  {L{\'o}pez-Sanjuan}, {D{\'{\i}}az-Garc{\'{\i}}a}, {Varela}, {Stefanon},
  {Cenarro}, {Matute}, {Masegosa}, {M{\'a}rquez}, {Perea}, {Del Olmo},
  {Husillos}, {Alfaro}, {Aparicio-Villegas}, {Cervi{\~n}o}, {Huertas-Company},
  {Aguerri}, {Broadhurst}, {Cabrera-Ca{\~n}o}, {Cepa}, {Gonz{\'a}lez},
  {Infante}, {Mart{\'{\i}}nez}, {Prada}, \& {Quintana}}]{molino}
{Molino}, A., {Ben{\'{\i}}tez}, N., {Moles}, M., {et~al.} 2014, \mnras, 441,
  2891

\bibitem[{{Molino} {et~al.}(2018){Molino}, {Costa-Duarte}, {Mendes de
  Oliveira}, \& {J-PLUS collaboration}}]{molino18}
{Molino}, A., {Costa-Duarte}, M.~V., {Mendes de Oliveira}, C., \& {J-PLUS
  collaboration}. 2018, \aap, in press [arXiv:1804.03640]

\bibitem[{{Oke} \& {Gunn}(1983)}]{AB}
{Oke}, J.~B. \& {Gunn}, J.~E. 1983, \apj, 266, 713

\bibitem[{{Patton} {et~al.}(2011){Patton}, {Ellison}, {Simard}, {McConnachie},
  \& {Mendel}}]{patton}
{Patton}, D.~R., {Ellison}, S.~L., {Simard}, L., {McConnachie}, A.~W., \&
  {Mendel}, J.~T. 2011, \mnras, 412, 591

\bibitem[{{Roth} {et~al.}(2005){Roth}, {Kelz}, {Fechner}, {Hahn}, {Bauer},
  {Becker}, {B{\"o}hm}, {Christensen}, {Dionies}, {Paschke}, {Popow}, {Wolter},
  {Schmoll}, {Laux}, \& {Altmann}}]{PPAK}
{Roth}, M.~M., {Kelz}, A., {Fechner}, T., {et~al.} 2005, \pasp, 117, 620

\bibitem[{{Salpeter}(1955)}]{salpeter}
{Salpeter}, E.~E. 1955, \apj, 121, 161

\bibitem[{{San Roman} {et~al.}(2018){San Roman}, {S{\'a}nchez-Bl{\'a}zquez},
  {Cenarro}, \& {J-PLUS collaboration}}]{sanroman18}
{San Roman}, I., {S{\'a}nchez-Bl{\'a}zquez}, P., {Cenarro}, A.~J., \& {J-PLUS
  collaboration}. 2018, \aap, in prep.

\bibitem[{{Sanchez} {et~al.}(2017){Sanchez}, {Avila-Reese}, {Hernandez-Toledo},
  {Cortes-Suarez}, {Rodriguez-Puebla}, {Ibarra-Medel}, {Cano-Diaz}, {Negrete},
  {Calette}, {de Lorenzo-Caceres}, {Ortega-Minakata}, {Aquino}, {Valenzuela},
  {Clemente}, {Storchi-Bergmann}, {Riffel}, {Schimoia}, {Riffel}, {Rembold},
  {Brownstein}, {Pan}, {Robes}, \& {Mallmann}}]{sanchezAGN}
{Sanchez}, S.~F., {Avila-Reese}, V., {Hernandez-Toledo}, H., {et~al.} 2017,
  ArXiv e-prints [\eprint[arXiv]{1709.05438}]

\bibitem[{{S{\'a}nchez} {et~al.}(2016{\natexlab{a}}){S{\'a}nchez},
  {Garc{\'{\i}}a-Benito}, {Zibetti}, {Walcher}, {Husemann}, {Mendoza},
  {Galbany}, {Falc{\'o}n-Barroso}, {Mast}, {Aceituno}, {Aguerri}, {Alves},
  {Amorim}, {Ascasibar}, {Barrado-Navascues}, {Barrera-Ballesteros},
  {Bekerait{\`e}}, {Bland-Hawthorn}, {Cano D{\'{\i}}az}, {Cid Fernandes},
  {Cavichia}, {Cortijo}, {Dannerbauer}, {Demleitner}, {D{\'{\i}}az}, {Dettmar},
  {de Lorenzo-C{\'a}ceres}, {del Olmo}, {Galazzi}, {Garc{\'{\i}}a-Lorenzo},
  {Gil de Paz}, {Gonz{\'a}lez Delgado}, {Holmes}, {Igl{\'e}sias-P{\'a}ramo},
  {Kehrig}, {Kelz}, {Kennicutt}, {Kleemann}, {Lacerda}, {L{\'o}pez
  Fern{\'a}ndez}, {L{\'o}pez S{\'a}nchez}, {Lyubenova}, {Marino},
  {M{\'a}rquez}, {Mendez-Abreu}, {Moll{\'a}}, {Monreal-Ibero}, {Ortega
  Minakata}, {Torres-Papaqui}, {P{\'e}rez}, {Rosales-Ortega}, {Roth},
  {S{\'a}nchez-Bl{\'a}zquez}, {Schilling}, {Spekkens}, {Vale Asari}, {van den
  Bosch}, {van de Ven}, {Vilchez}, {Wild}, {Wisotzki}, {Y{\i}ld{\i}r{\i}m}, \&
  {Ziegler}}]{califadr3}
{S{\'a}nchez}, S.~F., {Garc{\'{\i}}a-Benito}, R., {Zibetti}, S., {et~al.}
  2016{\natexlab{a}}, \aap, 594, A36

\bibitem[{{S{\'a}nchez} {et~al.}(2006){S{\'a}nchez}, {Garc{\'{\i}}a-Lorenzo},
  {Jahnke}, {Mediavilla}, {Gonz{\'a}lez-Serrano}, {Christensen}, \&
  {Wisotzki}}]{fit3D1}
{S{\'a}nchez}, S.~F., {Garc{\'{\i}}a-Lorenzo}, B., {Jahnke}, K., {et~al.} 2006,
  \nar, 49, 501

\bibitem[{{S{\'a}nchez} {et~al.}(2012{\natexlab{a}}){S{\'a}nchez}, {Kennicutt},
  {Gil de Paz}, {van de Ven}, {V{\'{\i}}lchez}, {Wisotzki}, {Walcher}, {Mast},
  {Aguerri}, {Albiol-P{\'e}rez}, {Alonso-Herrero}, {Alves}, {Bakos},
  {Bart{\'a}kov{\'a}}, {Bland-Hawthorn}, {Boselli}, {Bomans},
  {Castillo-Morales}, {Cortijo-Ferrero}, {de Lorenzo-C{\'a}ceres}, {Del Olmo},
  {Dettmar}, {D{\'{\i}}az}, {Ellis}, {Falc{\'o}n-Barroso}, {Flores},
  {Gallazzi}, {Garc{\'{\i}}a-Lorenzo}, {Gonz{\'a}lez Delgado}, {Gruel},
  {Haines}, {Hao}, {Husemann}, {Igl{\'e}sias-P{\'a}ramo}, {Jahnke}, {Johnson},
  {Jungwiert}, {Kalinova}, {Kehrig}, {Kupko}, {L{\'o}pez-S{\'a}nchez},
  {Lyubenova}, {Marino}, {M{\'a}rmol-Queralt{\'o}}, {M{\'a}rquez}, {Masegosa},
  {Meidt}, {Mendez-Abreu}, {Monreal-Ibero}, {Montijo}, {Mour{\~a}o},
  {Palacios-Navarro}, {Papaderos}, {Pasquali}, {Peletier}, {P{\'e}rez},
  {P{\'e}rez}, {Quirrenbach}, {Rela{\~n}o}, {Rosales-Ortega}, {Roth},
  {Ruiz-Lara}, {S{\'a}nchez-Bl{\'a}zquez}, {Sengupta}, {Singh}, {Stanishev},
  {Trager}, {Vazdekis}, {Viironen}, {Wild}, {Zibetti}, \& {Ziegler}}]{CALIFA}
{S{\'a}nchez}, S.~F., {Kennicutt}, R.~C., {Gil de Paz}, A., {et~al.}
  2012{\natexlab{a}}, \aap, 538, A8

\bibitem[{{S{\'a}nchez} {et~al.}(2016{\natexlab{b}}){S{\'a}nchez}, {P{\'e}rez},
  {S{\'a}nchez-Bl{\'a}zquez}, {Garc{\'{\i}}a-Benito}, {Ibarra-Mede},
  {Gonz{\'a}lez}, {Rosales-Ortega}, {S{\'a}nchez-Menguiano}, {Ascasibar},
  {Bitsakis}, {Law}, {Cano-D{\'{\i}}az}, {L{\'o}pez-Cob{\'a}}, {Marino}, {Gil
  de Paz}, {L{\'o}pez-S{\'a}nchez}, {Barrera-Ballesteros}, {Galbany}, {Mast},
  {Abril-Melgarejo}, \& {Roman-Lopes}}]{Sanchez2}
{S{\'a}nchez}, S.~F., {P{\'e}rez}, E., {S{\'a}nchez-Bl{\'a}zquez}, P., {et~al.}
  2016{\natexlab{b}}, \rmxaa, 52, 171

\bibitem[{{S{\'a}nchez} {et~al.}(2016{\natexlab{c}}){S{\'a}nchez}, {P{\'e}rez},
  {S{\'a}nchez-Bl{\'a}zquez}, {Gonz{\'a}lez}, {Ros{\'a}lez-Ortega},
  {Cano-D{\'{\i}} az}, {L{\'o}pez-Cob{\'a}}, {Marino}, {Gil de Paz},
  {Moll{\'a}}, {L{\'o}pez-S{\'a}nchez}, {Ascasibar}, \&
  {Barrera-Ballesteros}}]{Sanchez1}
{S{\'a}nchez}, S.~F., {P{\'e}rez}, E., {S{\'a}nchez-Bl{\'a}zquez}, P., {et~al.}
  2016{\natexlab{c}}, \rmxaa, 52, 21

\bibitem[{{S{\'a}nchez} {et~al.}(2011){S{\'a}nchez}, {Rosales-Ortega},
  {Kennicutt}, {Johnson}, {Diaz}, {Pasquali}, \& {Hao}}]{fit3D2}
{S{\'a}nchez}, S.~F., {Rosales-Ortega}, F.~F., {Kennicutt}, R.~C., {et~al.}
  2011, \mnras, 410, 313

\bibitem[{{S{\'a}nchez} {et~al.}(2012{\natexlab{b}}){S{\'a}nchez},
  {Rosales-Ortega}, {Marino}, {Iglesias-P{\'a}ramo}, {V{\'{\i}}lchez},
  {Kennicutt}, {D{\'{\i}}az}, {Mast}, {Monreal-Ibero}, {Garc{\'{\i}}a-Benito},
  {Bland-Hawthorn}, {P{\'e}rez}, {Gonz{\'a}lez Delgado}, {Husemann},
  {L{\'o}pez-S{\'a}nchez}, {Cid Fernandes}, {Kehrig}, {Walcher}, {Gil de Paz},
  \& {Ellis}}]{sanchezhii}
{S{\'a}nchez}, S.~F., {Rosales-Ortega}, F.~F., {Marino}, R.~A., {et~al.}
  2012{\natexlab{b}}, \aap, 546, A2

\bibitem[{{S{\'a}nchez-Bl{\'a}zquez} {et~al.}(2006){S{\'a}nchez-Bl{\'a}zquez},
  {Peletier}, {Jim{\'e}nez-Vicente}, {Cardiel}, {Cenarro},
  {Falc{\'o}n-Barroso}, {Gorgas}, {Selam}, \& {Vazdekis}}]{Patricia}
{S{\'a}nchez-Bl{\'a}zquez}, P., {Peletier}, R.~F., {Jim{\'e}nez-Vicente}, J.,
  {et~al.} 2006, \mnras, 371, 703

\bibitem[{{Sarzi} {et~al.}(2006){Sarzi}, {Falc{\'o}n-Barroso}, {Davies},
  {Bacon}, {Bureau}, {Cappellari}, {de Zeeuw}, {Emsellem}, {Fathi},
  {Krajnovi{\'c}}, {Kuntschner}, {McDermid}, \& {Peletier}}]{sarzi}
{Sarzi}, M., {Falc{\'o}n-Barroso}, J., {Davies}, R.~L., {et~al.} 2006, \mnras,
  366, 1151

\bibitem[{{Schaefer} {et~al.}(2017){Schaefer}, {Croom}, {Allen}, {Brough},
  {Medling}, {Ho}, {Scott}, {Richards}, {Pracy}, {Gunawardhana}, {Norberg},
  {Alpaslan}, {Bauer}, {Bekki}, {Bland-Hawthorn}, {Bloom}, {Bryant}, {Couch},
  {Driver}, {Fogarty}, {Foster}, {Goldstein}, {Green}, {Hopkins},
  {Konstantopoulos}, {Lawrence}, {L{\'o}pez-S{\'a}nchez}, {Lorente}, {Owers},
  {Sharp}, {Sweet}, {Taylor}, {van de Sande}, {Walcher}, \& {Wong}}]{schaefer}
{Schaefer}, A.~L., {Croom}, S.~M., {Allen}, J.~T., {et~al.} 2017, \mnras, 464,
  121

\bibitem[{{Scudder} {et~al.}(2012){Scudder}, {Ellison}, {Torrey}, {Patton}, \&
  {Mendel}}]{scuder}
{Scudder}, J.~M., {Ellison}, S.~L., {Torrey}, P., {Patton}, D.~R., \& {Mendel},
  J.~T. 2012, \mnras, 426, 549

\bibitem[{{Taylor} {et~al.}(2011){Taylor}, {Hopkins}, {Baldry}, {Brown},
  {Driver}, {Kelvin}, {Hill}, {Robotham}, {Bland-Hawthorn}, {Jones}, {Sharp},
  {Thomas}, {Liske}, {Loveday}, {Norberg}, {Peacock}, {Bamford}, {Brough},
  {Colless}, {Cameron}, {Conselice}, {Croom}, {Frenk}, {Gunawardhana},
  {Kuijken}, {Nichol}, {Parkinson}, {Phillipps}, {Pimbblet}, {Popescu},
  {Prescott}, {Sutherland}, {Tuffs}, {van Kampen}, \& {Wijesinghe}}]{masas}
{Taylor}, E.~N., {Hopkins}, A.~M., {Baldry}, I.~K., {et~al.} 2011, \mnras, 418,
  1587

\bibitem[{{Thomas} {et~al.}(2013){Thomas}, {Steele}, {Maraston}, {Johansson},
  {Beifiori}, {Pforr}, {Str{\"o}mb{\"a}ck}, {Tremonti}, {Wake}, {Bizyaev},
  {Bolton}, {Brewington}, {Brownstein}, {Comparat}, {Kneib}, {Malanushenko},
  {Malanushenko}, {Oravetz}, {Pan}, {Parejko}, {Schneider}, {Shelden},
  {Simmons}, {Snedden}, {Tanaka}, {Weaver}, \& {Yan}}]{tomas}
{Thomas}, D., {Steele}, O., {Maraston}, C., {et~al.} 2013, \mnras, 431, 1383

\bibitem[{{Vazdekis} {et~al.}(2010){Vazdekis}, {S{\'a}nchez-Bl{\'a}zquez},
  {Falc{\'o}n-Barroso}, {Cenarro}, {Beasley}, {Cardiel}, {Gorgas}, \&
  {Peletier}}]{vazdekis}
{Vazdekis}, A., {S{\'a}nchez-Bl{\'a}zquez}, P., {Falc{\'o}n-Barroso}, J.,
  {et~al.} 2010, \mnras, 404, 1639

\bibitem[{{Vilella-Rojo} {et~al.}(2015){Vilella-Rojo}, {Viironen},
  {L{\'o}pez-Sanjuan}, {Cenarro}, {Varela}, {D{\'{\i}}az-Garc{\'{\i}}a},
  {Crist{\'o}bal-Hornillos}, {Ederoclite}, {Mar{\'{\i}}n-Franch}, \&
  {Moles}}]{Gonzalo15}
{Vilella-Rojo}, G., {Viironen}, K., {L{\'o}pez-Sanjuan}, C., {et~al.} 2015,
  \aap, 580, A47

\bibitem[{{Wake} {et~al.}(2017){Wake}, {Bundy}, {Diamond-Stanic}, {Yan},
  {Blanton}, {Bershady}, {S{\'a}nchez-Gallego}, {Drory}, {Jones}, {Kauffmann},
  {Law}, {Li}, {MacDonald}, {Masters}, {Thomas}, {Tinker}, {Weijmans}, \&
  {Brownstein}}]{wake}
{Wake}, D.~A., {Bundy}, K., {Diamond-Stanic}, A.~M., {et~al.} 2017, \aj, 154,
  86

\bibitem[{{Walcher} {et~al.}(2014){Walcher}, {Wisotzki}, {Bekerait{\'e}},
  {Husemann}, {Iglesias-P{\'a}ramo}, {Backsmann}, {Barrera Ballesteros},
  {Catal{\'a}n-Torrecilla}, {Cortijo}, {del Olmo}, {Garcia Lorenzo},
  {Falc{\'o}n-Barroso}, {Jilkova}, {Kalinova}, {Mast}, {Marino},
  {M{\'e}ndez-Abreu}, {Pasquali}, {S{\'a}nchez}, {Trager}, {Zibetti},
  {Aguerri}, {Alves}, {Bland-Hawthorn}, {Boselli}, {Castillo Morales}, {Cid
  Fernandes}, {Flores}, {Galbany}, {Gallazzi}, {Garc{\'{\i}}a-Benito}, {Gil de
  Paz}, {Gonz{\'a}lez-Delgado}, {Jahnke}, {Jungwiert}, {Kehrig}, {Lyubenova},
  {M{\'a}rquez Perez}, {Masegosa}, {Monreal Ibero}, {P{\'e}rez}, {Quirrenbach},
  {Rosales-Ortega}, {Roth}, {Sanchez-Blazquez}, {Spekkens}, {Tundo}, {van de
  Ven}, {Verheijen}, {Vilchez}, \& {Ziegler}}]{sample}
{Walcher}, C.~J., {Wisotzki}, L., {Bekerait{\'e}}, S., {et~al.} 2014, \aap,
  569, A1

\bibitem[{{Wenger} {et~al.}(2000){Wenger}, {Ochsenbein}, {Egret}, {Dubois},
  {Bonnarel}, {Borde}, {Genova}, {Jasniewicz}, {Lalo{\"e}}, {Lesteven}, \&
  {Monier}}]{simbad}
{Wenger}, M., {Ochsenbein}, F., {Egret}, D., {et~al.} 2000, \aaps, 143, 9

\bibitem[{{Woods} {et~al.}(2006){Woods}, {Geller}, \& {Barton}}]{redshifts}
{Woods}, D.~F., {Geller}, M.~J., \& {Barton}, E.~J. 2006, \aj, 132, 197

\bibitem[{{York} {et~al.}(2000){York}, {Adelman}, {Anderson}, {Anderson},
  {Annis}, {Bahcall}, {Bakken}, {Barkhouser}, {Bastian}, {Berman}, {Boroski},
  {Bracker}, {Briegel}, {Briggs}, {Brinkmann}, {Brunner}, {Burles}, {Carey},
  {Carr}, {Castander}, {Chen}, {Colestock}, {Connolly}, {Crocker}, {Csabai},
  {Czarapata}, {Davis}, {Doi}, {Dombeck}, {Eisenstein}, {Ellman}, {Elms},
  {Evans}, {Fan}, {Federwitz}, {Fiscelli}, {Friedman}, {Frieman}, {Fukugita},
  {Gillespie}, {Gunn}, {Gurbani}, {de Haas}, {Haldeman}, {Harris}, {Hayes},
  {Heckman}, {Hennessy}, {Hindsley}, {Holm}, {Holmgren}, {Huang}, {Hull},
  {Husby}, {Ichikawa}, {Ichikawa}, {Ivezi{\'c}}, {Kent}, {Kim}, {Kinney},
  {Klaene}, {Kleinman}, {Kleinman}, {Knapp}, {Korienek}, {Kron}, {Kunszt},
  {Lamb}, {Lee}, {Leger}, {Limmongkol}, {Lindenmeyer}, {Long}, {Loomis},
  {Loveday}, {Lucinio}, {Lupton}, {MacKinnon}, {Mannery}, {Mantsch}, {Margon},
  {McGehee}, {McKay}, {Meiksin}, {Merelli}, {Monet}, {Munn}, {Narayanan},
  {Nash}, {Neilsen}, {Neswold}, {Newberg}, {Nichol}, {Nicinski}, {Nonino},
  {Okada}, {Okamura}, {Ostriker}, {Owen}, {Pauls}, {Peoples}, {Peterson},
  {Petravick}, {Pier}, {Pope}, {Pordes}, {Prosapio}, {Rechenmacher}, {Quinn},
  {Richards}, {Richmond}, {Rivetta}, {Rockosi}, {Ruthmansdorfer}, {Sandford},
  {Schlegel}, {Schneider}, {Sekiguchi}, {Sergey}, {Shimasaku}, {Siegmund},
  {Smee}, {Smith}, {Snedden}, {Stone}, {Stoughton}, {Strauss}, {Stubbs},
  {SubbaRao}, {Szalay}, {Szapudi}, {Szokoly}, {Thakar}, {Tremonti}, {Tucker},
  {Uomoto}, {Vanden Berk}, {Vogeley}, {Waddell}, {Wang}, {Watanabe},
  {Weinberg}, {Yanny}, {Yasuda}, \& {SDSS Collaboration}}]{sloan}
{York}, D.~G., {Adelman}, J., {Anderson}, Jr., J.~E., {et~al.} 2000, \aj, 120,
  1579

\end{thebibliography}

\end{document}